\pdfoutput=1

\documentclass[pdflatex]{ptephy_v1}

\preprintnumber{XXXX-XXXX} 

\usepackage{lineno}

\newcommand\rev[1]{#1}
\usepackage{ulem}
\newcommand\response[1]{#1}            

\newcommand\ourmuonbeam{reaccelerated thermal muon beam }

\begin{document}

\title{A New Approach for Measuring the Muon Anomalous Magnetic Moment and Electric Dipole Moment}


\author{M.~Abe}
\affil[1]{High Energy Accelerator Research Organization (KEK), Ibaraki, Japan}
\author[2,3]{S.~Bae}
\affil[2]{Seoul National University, Seoul, Republic of Korea}
\affil[3]{Institute for Nuclear and Particle Astrophysics, Seoul, Republic of Korea}
\author[4]{G.~Beer}
\affil[4]{University of Victoria, British Columbia, Canada}
\author[5]{G.~Bunce}
\affil[5]{Retired, Boulder, Colorado, USA}
\author[2,3]{H.~Choi}
\author[2,3]{S.~Choi}
\author[6]{M.~Chung}
\affil[6]{UNIST, Ulsan, Republic of Korea}
\author[7]{W.~da~Silva}
\affil[7]{LPNHE (CNRS/IN2P3/UPMC/UDD), Paris, France}
\author[8,9,10]{S.~Eidelman}
\affil[8]{Budker Institute of Nuclear Physics, Novosibirsk, Russia}
\affil[9]{Novosibirsk State University, Novosibirsk, Russia}
\affil[10]{Lebedev Physical Institute RAS, Moscow, Russia}
\author[11]{M.~Finger}
\affil[11]{Charles University, Prague, Czech Republic}
\author[1]{Y.~Fukao}
\author[12]{T.~Fukuyama}
\affil[12]{Osaka University, Osaka, Japan}
\author[13]{S.~Haciomeroglu}
\affil[13]{Institute for Basic Science (IBS), Daejeon, Republic of Korea}
\author[14]{K.~Hasegawa}
\affil[14]{Japan Atomic Energy Agency (JAEA), Ibaraki, Japan}
\author[15]{K.~Hayasaka}
\affil[15]{Niigata University, Niigata, Japan}
\author[16]{N.~Hayashizaki}
\affil[16]{Tokyo Institute of Technology, Tokyo, Japan}
\author[1]{H.~Hisamatsu}
\author[17]{T.~Iijima}
\affil[17]{Nagoya University, Aichi, Japan}
\author[18]{H.~Iinuma}
\affil[18]{Ibaraki University, Ibaraki, Japan}
\author[17]{K.~Inami}
\author[19]{H.~Ikeda}
\affil[19]{Japan Aerospace Exploration Agency (JAXA), Tokyo, Japan}
\author[1]{M.~Ikeno}
\author[20]{K.~Ishida}
\affil[20]{RIKEN, Saitama, Japan}
\author[12]{T.~Itahashi}
\author[20]{M.~Iwasaki}
\author[21]{Y.~Iwashita}
\affil[21]{Kyoto University, Kyoto, Japan}
\author[22]{Y.~Iwata}
\affil[22]{National Institute of Radiological Sciences (NIRS), Chiba, Japan}
\author[1]{R.~Kadono}
\author[23]{S.~Kamal}
\affil[23]{University of British Columbia, British Columbia, Canada}
\author[1]{T.~Kamitani}
\author[20]{S.~Kanda}
\author[7]{F.~Kapusta}
\author[24]{K.~Kawagoe}
\affil[24]{Kyushu University, Fukuoka, Japan}
\author[1]{N.~Kawamura}
\author[14]{R.~Kitamura}
\author[2,3]{B.~Kim}
\author[25]{Y.~Kim}
\affil[25]{Korea Advanced Institute of Science and Technology (KAIST), Daejeon, Republic of Korea}
\author[1]{T.~Kishishita}
\author[2,3]{H.~Ko}
\author[1]{T.~Kohriki}
\author[14]{Y.~Kondo}
\author[1]{T.~Kume}
\author[13]{M.~J.~Lee}
\author[13]{S.~Lee}
\author[26]{W.~Lee}
\affil[26]{Korea University, Seoul, Republic of Korea}
\author[27]{G.~M.~Marshall}
\affil[27]{TRIUMF, British Columbia, Canada}
\author[28]{Y.~Matsuda}
\affil[28]{The University of Tokyo, Tokyo, Japan}
\author[1,29]{T.~Mibe}
\affil[29]{Graduate University for Advanced Studies (SOKENDAI), Ibaraki, Japan}
\author[1]{Y.~Miyake}
\author[1]{T.~Murakami}
\author[1]{K.~Nagamine}
\author[1]{H.~Nakayama}
\author[1]{S.~Nishimura}
\author[1]{D.~Nomura}
\author[1]{T.~Ogitsu}
\author[1]{S.~Ohsawa}
\author[1]{K.~Oide}
\author[1]{Y.~Oishi}
\author[20]{S.~Okada}
\author[4,27]{A.~Olin}
\author[25]{Z.~Omarov}
\author[1]{M.~Otani}
\author[8,9]{G.~Razuvaev}
\author[29]{A.~Rehman}
\author[1,30]{N.~Saito}
\affil[30]{J-PARC Center, Ibaraki, Japan}
\author[20]{N.~F.~Saito}
\author[1]{K.~Sasaki}
\author[1]{O.~Sasaki}
\author[1]{N.~Sato}
\author[1]{Y.~Sato}
\author[25]{Y.~K.~Semertzidis}
\author[1]{H.~Sendai}
\author[31]{Y.~Shatunov}
\affil[31]{Budker Institute of Nuclear Physics SB RAS, Novosibirsk, Russia}
\author[1]{K.~Shimomura}
\author[1]{M.~Shoji}
\author[9,31]{B.~Shwartz}
\author[1]{P.~Strasser}
\author[17]{Y.~Sue}
\author[24]{T.~Suehara}
\author[6]{C.~Sung}
\author[17]{K.~Suzuki}
\author[1]{T.~Takatomi}
\author[1]{M.~Tanaka}
\author[24]{J.~Tojo}
\author[24]{Y.~Tsutsumi}
\author[1]{T.~Uchida}
\author[1]{K.~Ueno}
\author[20]{S.~Wada}
\author[26]{E.~Won}
\author[1]{H.~Yamaguchi}
\author[24]{T.~Yamanaka}
\author[1]{A.~Yamamoto}
\author[1]{T.~Yamazaki}
\author[28]{H.~Yasuda}
\author[1]{M.~Yoshida}
\author[24]{T.~Yoshioka}







\begin{abstract}%
This paper introduces a new approach to measure the muon magnetic moment anomaly $a_{\mu} = (g-2)/2$, and the muon electric dipole moment (EDM) $d_{\mu}$ at the J-PARC muon facility.
The goal of \response{our} experiment is to measure $a_{\mu}$ and $d_{\mu}$ using an independent method with a factor of 10 lower muon momentum, and a factor of 20 smaller diameter storage-ring solenoid
compared with previous and ongoing muon $g-2$ experiments
with unprecedented quality of the storage magnetic field.  Additional significant differences from the present experimental method include a factor of 1,000 smaller transverse emittance of the 
muon beam (reaccelerated thermal muon beam), its efficient vertical injection into the solenoid, and tracking each decay positron from muon decay to obtain its momentum vector.  The precision goal for $a_{\mu}$ is statistical uncertainty of 450 part per billion (ppb), similar to the present experimental uncertainty, and a systematic uncertainty less than 70~ppb. The goal for EDM is a sensitivity of 
$1.5\times 10^{-21}~e\cdot\mbox{cm}$.


\end{abstract}

\subjectindex{C31}

\maketitle


\newcommand {\ee}        {e^+e^-}
\newcommand {\mumu}      {\mu^+\mu^-}
\newcommand {\tautau}    {\tau^+\tau^-}
\newcommand {\ellell}    {\ell^+\ell^-}
\newcommand {\ra}        {\rightarrow}
\newcommand{\gsim}{\;\raisebox{-0.9ex}
           {$\textstyle\stackrel{\textstyle >}{\sim}$}\;}
\newcommand{\lsim}{\;\raisebox{-0.9ex}{$\textstyle\stackrel{\textstyle<}
           {\sim}$}\;}

\section{\rev{Introduction}}\label{sec:Introduction} 
%
The Standard Model (SM)~\cite{Weinberg:1967tq, Salam:1968rm} 
is an extremely successful theory of elementary particles.
Even though more than 50 years have passed since it was
first proposed, it \rev{remains}
the best effective theory which
can describe physics below the weak scale.
In fact, the recent discovery of the Higgs
boson~\cite{Aad:2012tfa, Chatrchyan:2012ufa} and
the measurements of its properties such as the signal
strengths at the \response{Large Hadron Collider} (LHC)~\cite{PDG}
have made our confidence in the SM stronger than ever.

Although the SM is such a successful theory, 
it is a firm expectation of many physicists that
the SM is not the ultimate theory to describe physics
at the shortest length scale.  There are a number of
reasons behind this.
Firstly, there are as many as 19 free parameters in the
SM whose values cannot be predicted from theory alone but
can be determined only by experiments.  Secondly, the SM
must somehow be extended to accommodate gravity.  It is known
that this is difficult, and one may need a 
much larger framework such as string theory.
Thirdly, in the SM, there is the gauge hierarchy problem, to explain why there are two
vastly different
fundamental scales, the weak scale 
$M_{\text{weak}} ~(= {\cal O}(100) {\text{ GeV}}/c^2)$
and the Planck scale 
$M_{\text{Pl}} ~(= {\cal O}(10^{18}) {\text{ GeV}}/c^2)$.

Presently, \rev{many} experiments are ongoing to search for new
physics beyond the SM.
\rev{Among the most promising are experiments at} the LHC which directly probe physics
at the TeV scale.  
\rev{To date, new physics has not been discovered},
and a limit of $m_{\tilde{g}, \tilde{q}} \gsim 1$ TeV/$c^2$
has been obtained on the masses of \rev{gluinos and squarks}, for
example~\cite{PDG}.  

In view of this situation, the role played by precision
measurements is becoming more crucial.  
Even when direct searches for new physics
\rev{are limited in energy reach}, indirect searches like \rev{precision 
measurements} can become powerful probes of new physics.
Moreover, it is reported~\cite{KNT18, Davier:2017zfy,
Jegerlehner:2017gek, Benayoun:2015gxa, Blum:2018} that \rev{there is at present a more than}
3\,$\sigma$ discrepancy between the experimental
value of the muon's anomalous magnetic moment 
($a_{\mu}=(g-2)/2$, where $g$ is the Land\'{e} $g$-factor of the muon)~\cite{Bennett:2006fi} and
prediction for it.  In fact, the SM prediction quoted
in Ref.~\cite{PDG} is
\begin{align}
  a_\mu( \text{SM} ) 
  = (11 \, 659 \, 182.3 \pm 0.1 \pm 3.4 \pm 2.6) \times 10^{-10} ~,
 \label{eq:a_mu_SM}
\end{align}
where the uncertainties are from the electroweak, 
leading-order hadronic, and higher-order hadronic contributions,
respectively.  This value should be compared with the current
experimental value [5,11],
\begin{align}
  a_\mu( \text{exp} ) 
  = (11 \, 659 \, 209.1 \pm 5.4 \pm 3.3) \times 10^{-10} ~,
 \label{eq:a_mu_exp}
\end{align}
where the errors are the statistical and systematic uncertainties,
respectively.   The difference between Eqs.~(\ref{eq:a_mu_SM})
and (\ref{eq:a_mu_exp}) is
\begin{align}
  \Delta a_\mu \equiv a_\mu( \text{exp} ) - a_\mu( \text{SM} )
  = (26.8 \pm 7.6) \times 10^{-10} ~,
\end{align}
which means a $3.5 \sigma$ deviation.  
This deviation may be the result of physics beyond the SM.
This is a major motivation for new measurements of $a_\mu$.

The reported deviation of the muon anomaly from the SM has another important
implication. Since \rev{the contribution} from new particles such
as the smuon and the Kaluza-Klein excitations of the muon
\rev{may be} responsible for the deviation, it is natural to 
expect that effects from such new particles may also appear
in closely related processes such as the muon electric dipole
moment (EDM)~\cite{Crivellin:2018qmi}, $\mu \to e\gamma$ and $\mu$-$e$ conversion in nuclei
(see, e.g., Ref.~\cite{Paradisi:2016ntl} for a recent concise review).
It is therefore valuable to study the muon EDM ($d_{\mu}$), in addition to
the muon $g-2$.

\begin{table}[t]
\caption{\label{T:Comparison}Comparison of BNL-E821, FNAL-E989, and \response{our} experiment
}
\begin{tabular}{|l|ccccc|}
\hline
              & BNL-E821 &~~~& Fermilab-E989 &~~~& \response{Our} Experiment \\
\hline 
\hline 
Muon momentum & \multicolumn{3}{c}{3.09~GeV/$c$} &~~~& 300~MeV/$c$ \\
Lorentz $\gamma$ & \multicolumn{3}{c}{29.3} &~~~&  3 \\ 
Polarization  & \multicolumn{3}{c}{ 100\% } &~~~& 50\% \\
Storage field & \multicolumn{3}{c}{$B =1.45~$T} &~~~& $B=3.0$~T \\
Focusing field & \multicolumn{3}{c}{Electric quadrupole} &~~~& Very weak magnetic\\
Cyclotron period  & \multicolumn{3}{c}{ 149~ns } &~~~& 7.4~ns \\
Spin precession period & \multicolumn{3}{c}{ 4.37~$\mu$s } &~~~& 2.11~$\mu$s \\
Number of detected $e^+$ & 5.0$\times 10^9$ &~~& 1.6$\times 10^{11}$ &~~~& $5.7 \times 10^{11}$\\
Number of detected $e^-$ & 3.6$\times 10^9$ &~~& $-$ &~~~& $-$ \\
\hline 
$a_{\mu}$ precision (stat.) & 460~ppb &~~~& 100~ppb &~~~& 450~ppb \\
~~~~~~~~~~~~~~~~~(syst.) & 280~ppb &~~~& 100~ppb &~~~& $<$70~ppb \\
EDM precision (stat.) & $0.2\times 10^{-19}$~$e\cdot\mbox{cm}$ &~~~& --- &~~~& $1.5 \times 10^{-21}$~$e\cdot\mbox{cm}$ \\
~~~~~~~~~~~~~~~~~~~~ (syst.) & $0.9\times 10^{-19}$~$e\cdot\mbox{cm}$ &~~~& --- &~~~& $0.36 \times 10^{-21}$~$e\cdot\mbox{cm}$ \\
\hline      
\end{tabular}
\end{table} 

The current experimental result for $a_{\mu}$ is from the E821 experiment 
at Brookhaven National Laboratory \response{(BNL)}~\cite{Bennett:2006fi}, 
which used the ``magic gamma" approach with 100\% polarized 3~GeV/$c$ muons 
injected by an inflector magnet with 2--5\% efficiency into a 14-meter-diameter 
storage ring built with 360~degree superconducting coils, 
12 iron back-leg sectors and 36 iron pole sectors. 
With iron shims, a 1 part per million (ppm) field uniformity was achieved
averaged over the muon orbit, with local non-uniformity of up to 100~ppm. 
Electrostatic focusing was used in the ring, and decay positrons (and electrons) 
were observed with calorimetry.  A new measurement of $a_{\mu}$ is underway 
at Fermilab~\cite{Grange:2015fou}, using the BNL-E821 storage ring,
with a new muon accumulator ring and significant magnetic shimming improvements, 
with expected gain in statistical and systematic uncertainties.

\response{Our} experiment introduced here
is intended to measure $a_{\mu}$ and $d_{\mu}$ with a very different technique,
using a 300~MeV/$c$ \ourmuonbeam with 50\% polarization,
vertically injected into an \response{Magnetic Resonance Imaging} (MRI)-type solenoid storage ring 
with 1~ppm local magnetic field uniformity for the muon storage region
\response{with an orbit diameter of 66~cm.}

The vertical injection, invented for \response{our} experiment, 
will improve injection efficiency by more than an order of magnitude.
Very weak magnetic focusing will be used in the ring. 
Silicon strip detectors in the field will measure 
the momentum vector of the decay positrons. 

Table~\ref{T:Comparison} compares \response{our} experiment with the previous
experiment BNL-E821, and the current experiment Fermilab-E989.
The initial goal of \response{our} experiment is to reach the statistical uncertainty for $a_{\mu}$ 
of BNL-E821, with much smaller systematic uncertainties from sources different from the current method.
The muon EDM goal is a statistical sensitivity of $1.5\times 10^{-21}~e\cdot\mbox{cm}$
with a systematic uncertainty of $0.36\times 10^{-21}~e\cdot\mbox{cm}$,
\response{which is a factor of 60 improvement over the present measurement~\cite{Bennett:2008dy},
$d_\mu (\text{exp}) = ( 0.0 \pm 0.2\text{(stat.)} \pm  0.9\text{(syst.)}) \times 10^{-19}~e \cdot \text{cm}$.}


\section{Overview of the experiment}\label{sec:Principle} 

The experiment measures $a_{\mu}$ and $\eta$. They are defined by the relations
\begin{equation}
a_{\mu} = \frac{g-2}{2} ~~~~{\rm with} ~~~~
\vec{\mu}_{\mu} = g \left( \frac{e}{2m}
                  \right) \vec{s}, ~~~~
\vec{d}_{\mu} = \eta \left( \frac{e}{2mc}
                  \right) \vec{s},
\end{equation}
where $e, m$ and $\vec{s}$ are the electric charge, mass, and spin vector of the muon, respectively.
Here, $g$ is the Land\'e $g$-factor and $\eta$ is a corresponding factor for the EDM.
The experiment stores spin polarized $\mu^{+}$ in a magnet and 
the muons orbit in the uniform magnetic field. 
The spin of the muon precesses in the magnetic field. 
With the non-zero and positive value for $g-2$,
the muon spin direction rotates faster than the momentum.

The spin precession vector with respect to its momentum 
in a static magnetic field $\vec{B}$ and electric field $\vec{E}$
is given as~\cite{Thomas1927,Bargmann1959,Nelson1959,Khriplovich1998,Fukuyama2013,Fukuyama2018}
\begin{eqnarray} 
\vec{\omega} & = & \vec{\omega}_{a} + \vec{\omega}_{\eta} \\ 
             & = & 
  - \frac{e}{m}\left[a_{\mu} \vec{B} - 
   \left( a_{\mu}- \frac{1}{\gamma^2-1} \right)
   \frac{\vec{\beta} \times \vec{E}}{c}
  +\frac{\eta}{2} 
   \left(\vec{\beta} \times \vec{B}  + \frac{\vec{E}}{c}
   \right)
   \right].
\label{eq:omega_full}
\end{eqnarray}
Here $\vec{\omega}_{a}$ and $\vec{\omega}_{\eta}$ are precession \rev{vectors}
due to $g-2$ and EDM. $\vec{\beta}$ and $\gamma$ are the velocity and Lorentz factor of the muon,
respectively.

In the previous $g-2$ measurements, the energy of the muon
was chosen to cancel the term of $\vec{\beta} \times \vec{E}$,
which allowed for electrostatic focusing in the storage ring 
without affecting the muon spin precession to first order. 
A focusing field index of $n =$0.12--0.14 was used, 
which was necessary to contain the muons captured from pion decay. 
In this proposed experiment, we greatly reduce the focusing requirement 
in the storage ring by using a \ourmuonbeam with a factor of 1,000 smaller beam emittance.
Very weak magnetic focusing with a field index of $n \sim 10^{-4}$
is enough to store the muon beam, using no electric field for focusing.
Under this condition, Eq.~(\ref{eq:omega_full}) reduces to
\begin{equation}
\vec{\omega}  =  
  - \frac{e}{m}\left[a_{\mu} \vec{B} 
  +\frac{\eta}{2} 
   \left(\vec{\beta} \times \vec{B} \right) 
   \right].
\label{eq:omega_tot}
\end{equation}
There is no contribution from the $\vec{\beta} \times \vec{E}$ term at any beam energy.
Since the precession vectors $\vec{\omega}_{a}$ and $\vec{\omega}_{\eta}$ 
are orthogonal, the $g-2$ and EDM precessions 
\rev{can be measured simultaneously with}
an appropriate detector design.

The key requirement for this new approach is a muon beam with low emittance.
This \rev{can} be realized with a source of 
positive muons with thermal energy followed by
reacceleration, without increasing the transverse momentum spread. 
We note here that the stopping \rev{muons and their} reacceleration steps will \rev{also}
allow us to frequently reverse the muon spins by using static electromagnetic fields.
This \rev{feature} will be a powerful tool to study
rate-dependent systematics such as track reconstruction efficiency, and the effect of pile-up hits.

In the extraction of $a_{\mu}$ and $\eta$, 
the precession frequency $\vec{\omega}$ and the magnetic field $\vec{B}$ must
be measured. The quantity $\vec{\omega}$ is measured by detecting positrons from muon decays during the storage.
Like the other experiments that measure the muon anomalous
moment, this method exploits the correlation of muon spin
direction, or the polarization direction of the positive muon
beam, with the energy and direction of the $e^+$ emitted in decay of
the circulating stored muons~\cite{Konopinski:1959}. By selecting the most
energetic $e^+$, the rate of detection will show an oscillation in
time due to the precession of the muon spin with respect to its
momentum direction in the storage field.
Detectors \rev{located} radially inside the muon storage orbit will track the decay
$e^{+}$. \response{Our} experiment records the number of higher energy $e^{+}$ 
versus time in storage, as the muon spin precesses in the magnetic field.

The average magnetic field seen by the muons in the storage ring 
is measured by the Larmor precession frequency of a free proton ($\omega_p$).
This is obtained from a convolution of the magnetic field map
and the muon beam distribution measured by the experiment.

Assuming the EDM term is negligibly small compared with the $g-2$ term in Eq.~(\ref{eq:omega_tot}), 
$a_\mu$ is obtained from 
$\omega_a = \frac{e}{m} a_\mu B$.
By using $\omega_p$, one can rewrite this equation to
\begin{eqnarray}
a_\mu = \frac{R}{\lambda - R},
\label{eq:amu}
\end{eqnarray}
where $R = \omega_a /\omega_p$ and $\lambda = \mu_\mu/\mu_p$ is the muon-to-proton
magnetic moment ratio provided by separate experiments.
The precision of the direct measurement of $\lambda$ by muonium spectroscopy in the magnetic field
is 120~ppb~\cite{Liu:1999}. 
A new improved measurement of $\lambda$ is being prepared at J-PARC 
Materials and Life science experimental Facility (MLF)
in the same beamline~\cite{MuSEUM}.

\begin{figure}[t]
  \begin{center}
      \includegraphics[width=15cm,bb=0 0 1440 936]{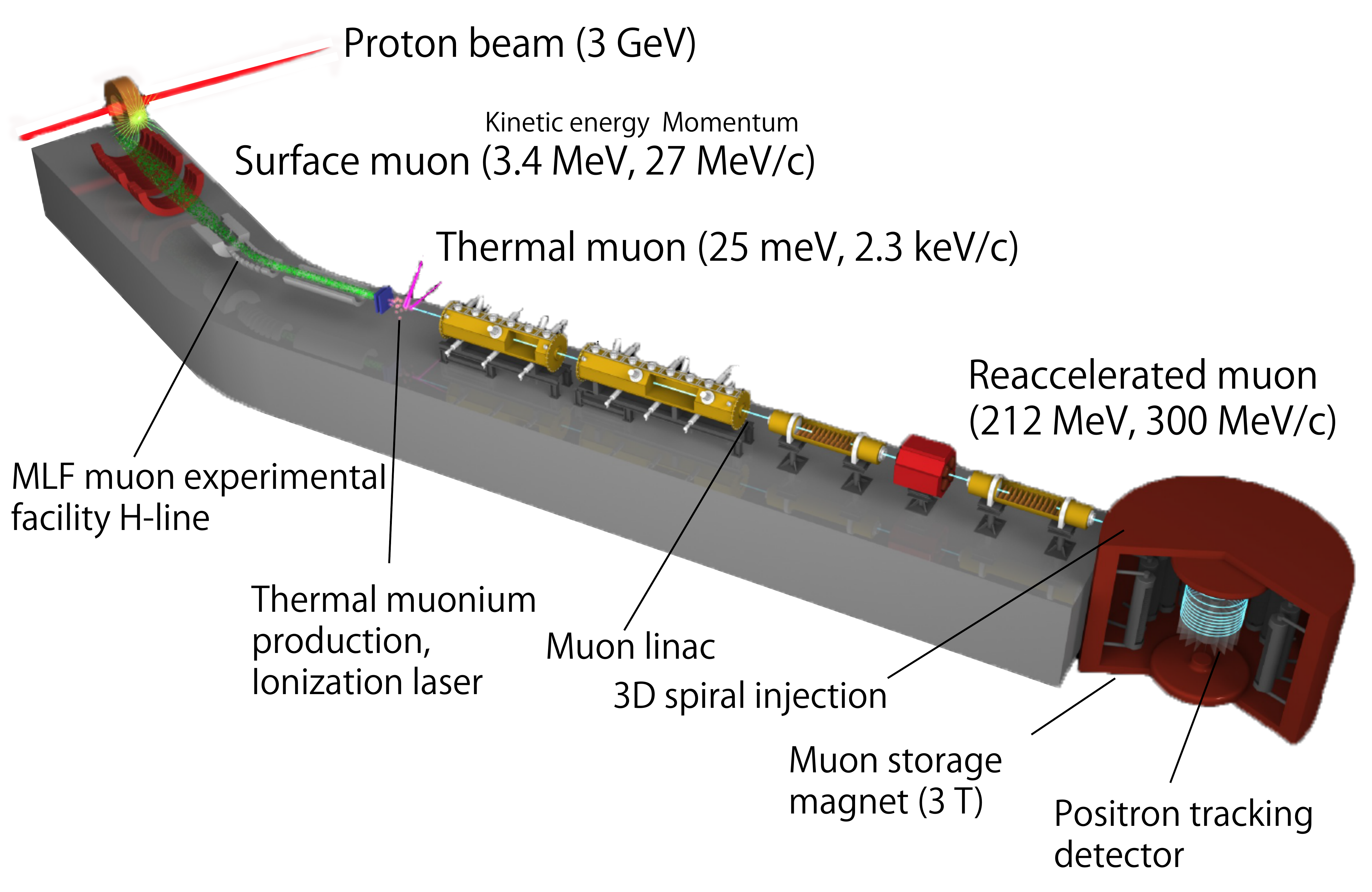}
  \end{center}
  \caption{Schematic view of the muon $g-2$/EDM experiment at J-PARC MLF. }
  \label{fig:setup}
\end{figure}

\response{Our} experiment will be installed at the muon facility (MUSE, Muon Science Establishment)~\cite{Kawamura:MUSE} in
the MLF of J-PARC.
A schematic of the experimental setup is shown in Fig.~\ref{fig:setup}. 
Experimental components and sensitivity estimations are described in the following sections.

\section{Experimental facility and surface muon beam}
A primary proton beam of 3~GeV kinetic energy with 1~MW beam power from the Rapid Cycle Synchrotron hits 
a 2~cm thick graphite target to provide pulsed muon beams.
The proton beam has a double-pulse structure, and each pulse is 100~ns in width (FWHM) 
with a 600~ns separation and 25~Hz repetition rate.
\response{Our} experiment uses a surface muon beam.
Surface muons are nearly 100\% polarized positive muons from the decay of pions 
stopped at and near the target surface with the consequent momentum of 29.8 MeV/$c$ and below.
There are four beamlines extracting muon beams.
\response{Our} experiment will use one of those, the H-line.

\begin{figure}[t]
  \begin{center}
    \includegraphics[bb=0  0 595 400, width=0.8\textwidth]{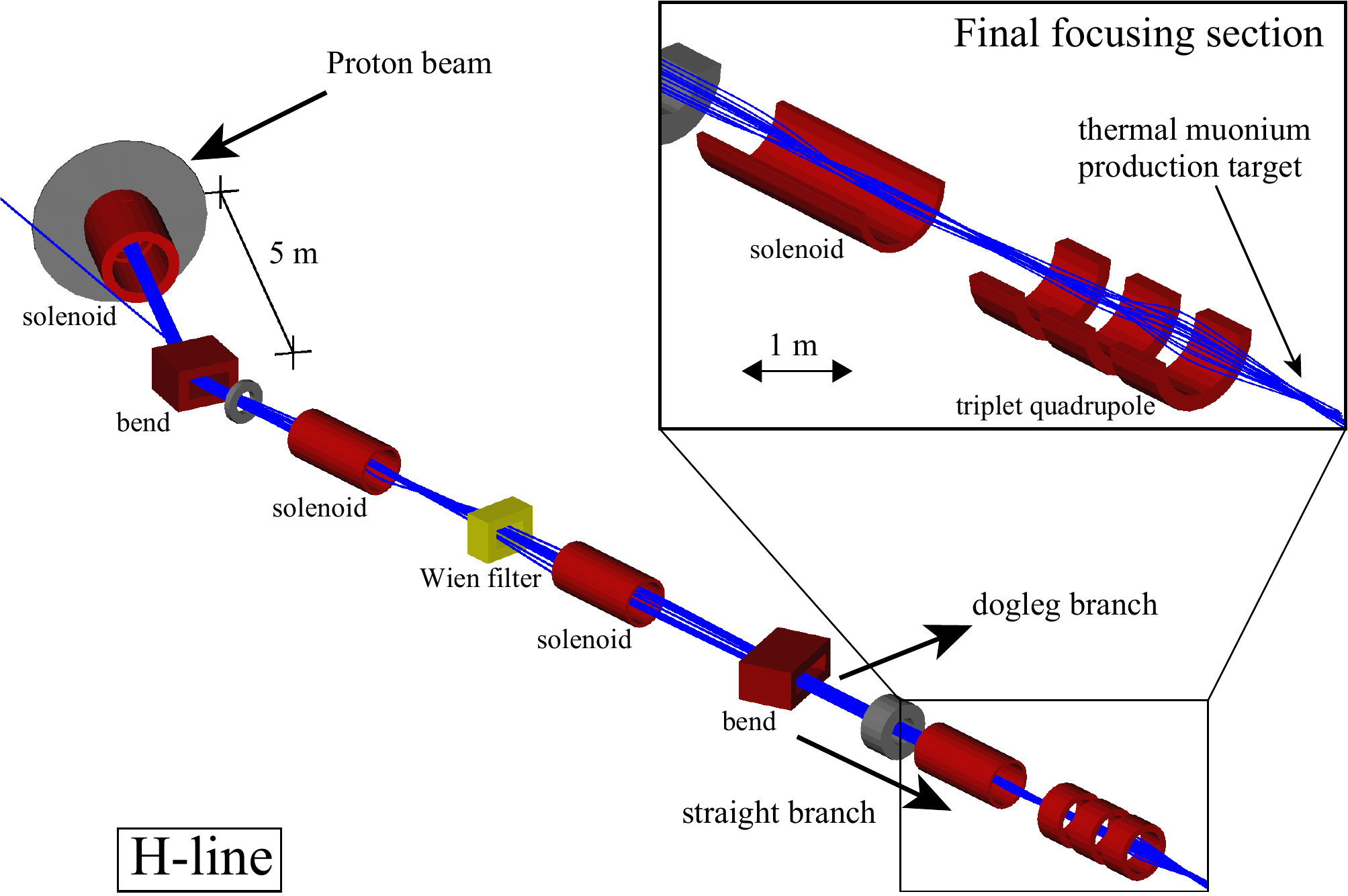}
  \end{center}
  \caption{
    Layout of the muon beamline (H-line) providing surface muons. Lines are simulated muon beam trajectories.
  }
  \label{fig:g4beamline_Hlineff}
\end{figure}

The H-line is a new beamline designed 
to deliver a high intensity muon beam~\cite{Kawamura:H-line}.
This is realized by 
adopting a large aperture solenoid magnet to capture muons from the muon production target,
wide gap bending magnets for momentum selection,
and a pair of opposite directional solenoid magnets for efficient beam transport.
The surface muon beam is focused onto a target to produce muonium atoms.
The final focus condition is optimized to maximize the number of muons stopping in the muonium production 
target and to minimize the leakage magnetic field at the focal point. To fulfill these 
requirements, 
the final focusing includes a solenoid magnet
followed by a triplet of quadrupole magnets.
The layout of the H-line is shown in Fig.~\ref{fig:g4beamline_Hlineff}.

The intensity of the surface muon beam at H-line is estimated 
to be $\sim 10^8$ per second at the designed proton beam power of 1 MW.
The surface muon at the end of the beamline has a momentum centered at $p =$ 27~MeV/$c$ with 
momentum spread \response{$\Delta p/p= $5\% (RMS)}. 
According to a beam transport simulation~\cite{Kawamura:G4bl}, the beam will be focused on the focal point with
the standard deviation of 31 and 14 mm to the horizontal and vertical direction, respectively.

\section{Production of thermal muons from surface muons} 
\begin{figure}[t]
\centering
\includegraphics[width=1.0\columnwidth]{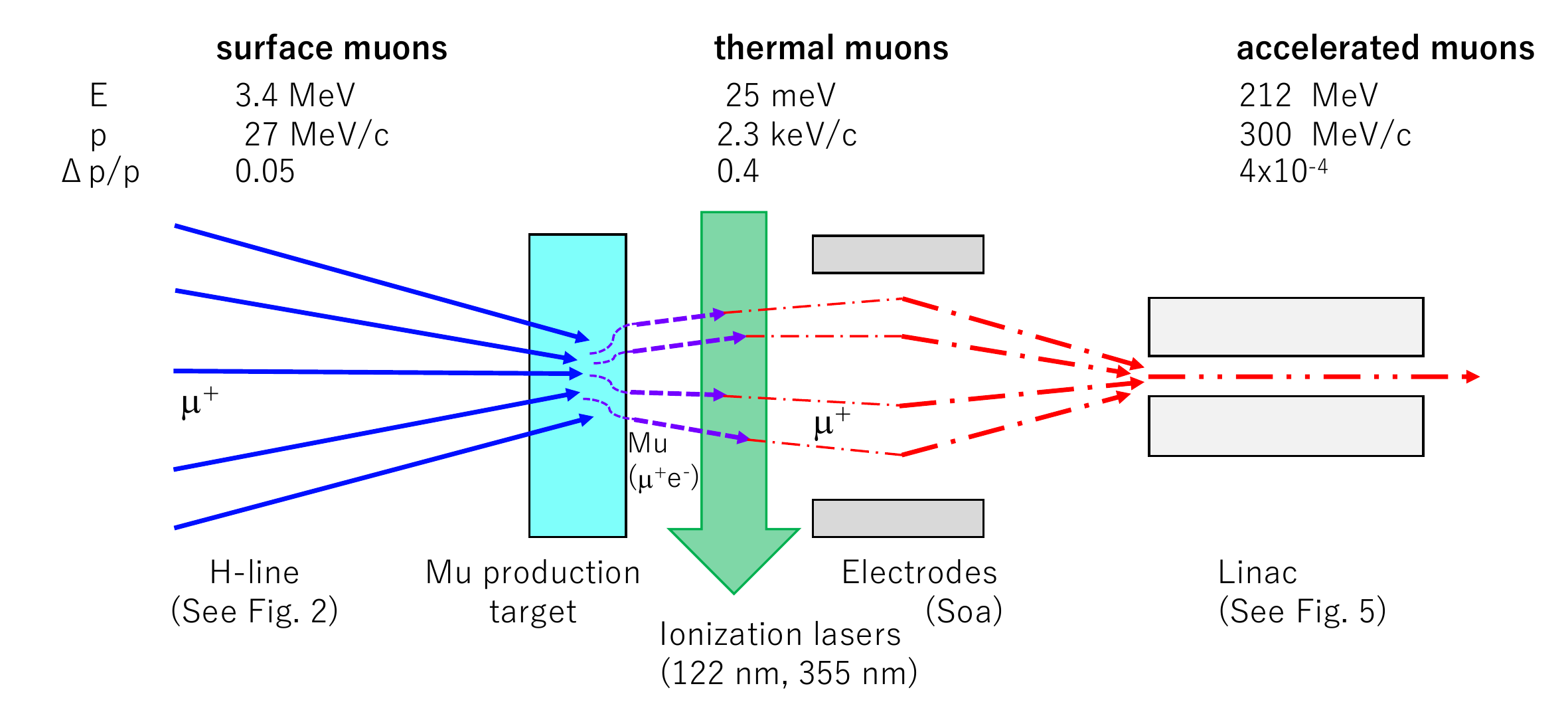}
\caption{
Scheme of the reaccelerated thermal muon beam. The surface
muon beam is thermalized in silica aerogel near the downstream
edge of the Mu production target slab. Some of the muonium formed
will diffuse to the surface of the slab and escape to vacuum with
thermal energy. Intense laser beams strip the electron from
muonium and the muon is accelerated by a static electric field
followed by RF linac structures.
Kinetic energy ($E$), total momentum ($p$), and its spread ($\Delta p/p$) 
at three stages are given.
}
\label{fig:usm} 
\end{figure}

The surface muon beam is converted at its final focus into a
source of room-temperature muons. The first step is to
slow down and thermalize the $\mu^{+}$ in a carefully selected
material, silica aerogel \cite{tabata:2012}.  In this
material, most of the muons form muonium atoms ($\mu^{+}e^{-}$, or Mu)~\cite{bakule:2013}
that diffuse as neutral atoms into a vacuum region where Mu is
ionized by laser excitation (Fig.~\ref{fig:usm}). While the thermalization, conversion
to Mu, diffusion, and ionization steps result in the loss of a significant
fraction of the original surface muon beam, the
characteristics of thermal muons after muonium ionization can be
exploited as a source for acceleration and injection into a
storage ring. A comparison of the kinematic characteristics of
surface muons, a thermal source, and accelerated muons
is summarized in Fig.~\ref{fig:usm}.


Very low density silica aerogel is chosen as the muonium production target for
high Mu formation probability ($>0.5$) and low
relaxation of the polarization.
The maximum polarization is 50\% after the statistical spin
distribution among hyperfine states settles in the Mu atom.
In addition, the silica aerogel provides a large mobility of
Mu atoms within the aerogel structure such that they can be
emitted with a near-thermal room temperature energy distribution
from the surface of the aerogel slab into the adjacent vacuum region.

The emission of Mu from aerogel, as well as the other important
characteristics described above, has been discovered and verified by experiments
on surface muon beam lines at TRIUMF~\cite{bakule:2013,beer:2014} and J-PARC.
The results showed that the
emission probability was enhanced by an order of magnitude if the
downstream aerogel surface was covered with a close-packed array of holes
produced by laser ablation to a depth of the order a few mm. 
The data are consistent with the assumption of Mu diffusion 
within the aerogel slab to the surface of the ablation holes 
followed by emission through the holes 
with speeds corresponding to thermal velocity near
room temperature.

\begin{figure}[t]
\centering
\includegraphics[width=0.8\columnwidth]{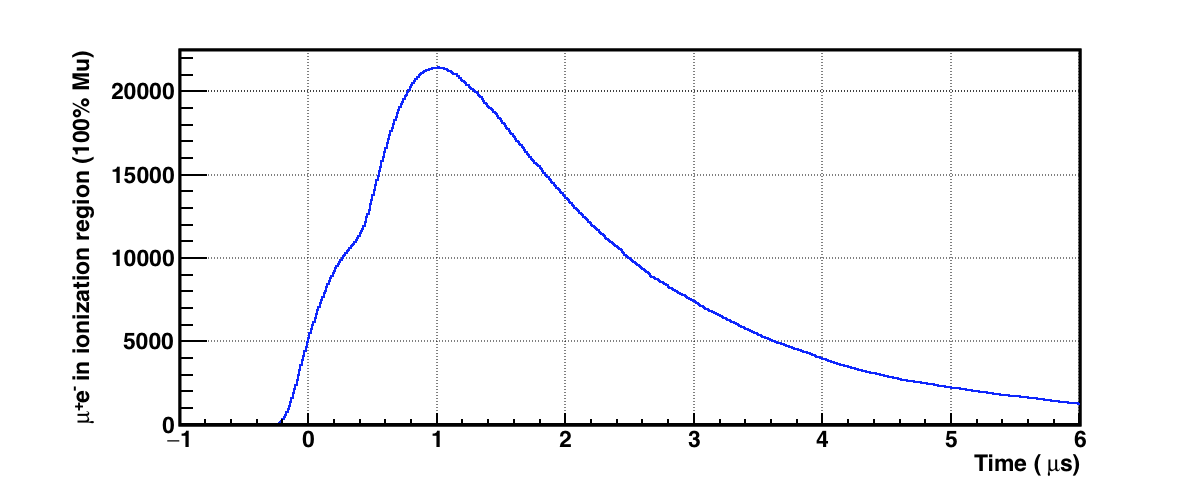}
\caption{Evolution of muonium into the laser irradiation region following diffusion and emission
from a laser-ablated aerogel target. This is the result of a diffusion simulation with parameters
that fit the results of Ref.~\cite{beer:2014}. The time origin is set at the middle of the double-pulse
structure of the surface muons.
The graph corresponds to the number of beam muons $3.23 \times 10^6$ and assumes 
100\% Mu formation per stopping muon. 
\response{We expect the probability of Mu in the laser irradiation region to be 0.0034 ($= 0.52 \times 2.1 \times 10^4 /(3.23 \times 10^6)$),  
where 0.52 is the initial formation probability of monism in the aerogel.}
}
\label{fig:t_Mu_in_laser} 
\end{figure}

Figure~\ref{fig:t_Mu_in_laser} shows the simulated evolution of muonium into the laser irradiation region
located at 1~mm from the surface of the aerogel slab.
\response{Here the simulation was performed using the diffusion model as explained above, 
where the diffusion parameter was predetermined so as to best describe the TRIUMF data~\cite{beer:2014}.}
The laser irradiation region is defined as a volume of $50\times 200\times 5$~mm$^3$ in the transverse directions
and the longitudinal direction, respectively.
This simulation indicates that the optimum time for the
short ionization pulse is near 1.0 $\mu$s after the
average time of arrival of the two surface muon pulses (0.6~$\mu$s apart).
The efficiency for thermal muonium production 
is estimated to be $3.4\times 10^{-3}$ per surface muon.

A high-power ionizing laser system is synchronized to the
periodic 25~Hz thermal Mu production at its maximum density in vacuum.
The laser ionization consists of two processes. 
The first is $1s \rightarrow 2p$ excitation by a beam having the wavelength of 122~nm (Lyman-$\alpha$), 
and the second is electron dissociation by a \response{laser} beam with the wavelength of 355~nm. 
The spectral linewidth and the pulse energy of the excitation  beam is 80~GHz and 100~$\mu$J, respectively. 
The pulse energy of the ionization beam is 440~mJ. The pulse width of each beam is 1~ns. 
\response{The ionization efficiency was calculated to be 73\% based on the transition rates given by theoretical excitation and dissociation cross sections 
multiplied with the expected laser photon density.}
The coherent Lyman-$\alpha$ light is generated by a non-linear conversion in Kr gas from two pump laser beams.
Two pump beams for the frequency conversion are generated by a distributed feedback laser followed by four stages of amplifiers and three stages of frequency converters with nonlinear optical crystals.
Such an intense Lyman-$\alpha$ laser \cite{laser:2016} is being developed 
in collaboration with the group developing an ultra slow muon microscope, which 
is being used for the ionization of muonium at J-PARC U-line \cite{usm:2018}.

\section{Acceleration}

\begin{figure}[t]
\centering
\includegraphics*[width=1.0\textwidth]{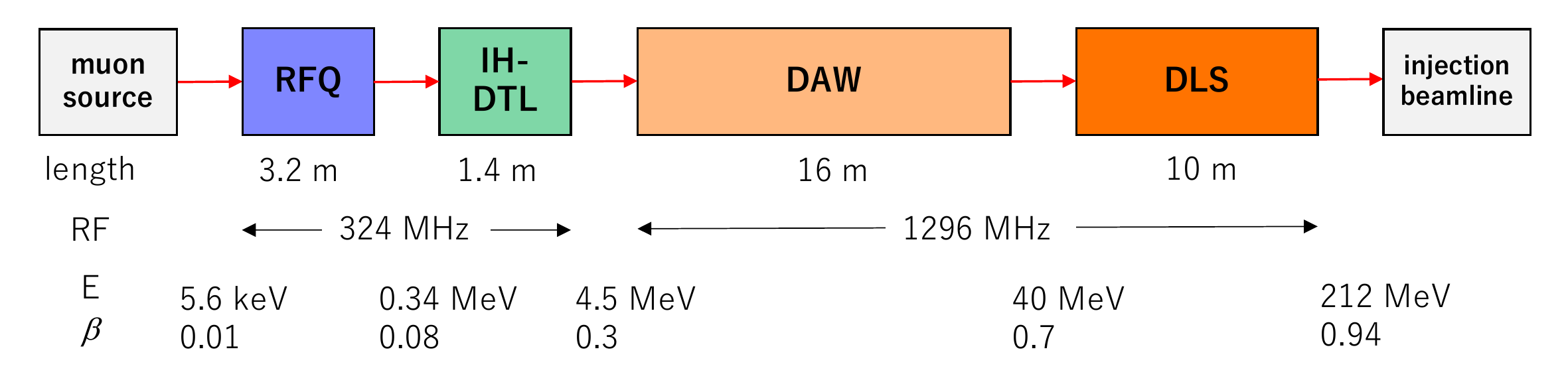}
\caption{Schematic configuration of the muon linac.}
\label{fig:mulinac_config}
\end{figure}

The room-temperature muons created by the laser ionization of thermal muonium will be 
accelerated to a momentum of 300~MeV/$c$ (212~MeV in kinetic energy).
The muons must be accelerated in a sufficiently short time
compared with the muon lifetime of 2.2~$\mu$s
to suppress muon decay loss during the acceleration.
Another essential requirement for the acceleration is
the suppression of transverse emittance growth.
To satisfy these, a linac dedicated to this purpose will be used in \response{our} experiment.
Figure~\ref{fig:mulinac_config} shows the schematic configuration
of the muon linac.
In accelerating the muons, the $\beta$ increases rapidly with the
kinetic energy.
It is important to adopt adequate 
accelerating structures to obtain high acceleration efficiency,
similar to proton linacs. 
The acceleration steps are 1) electrostatic acceleration with a Soa lens,
2) radio frequency quadrupole (RFQ),
3) interdigital H-type drift tube linac (IH-DTL),
4) disk-and-washer structure (DAW),
and 5) disk-loaded traveling wave structure (DLS).


\begin{figure}[t]
\centering
\includegraphics*[width=1.01\textwidth]{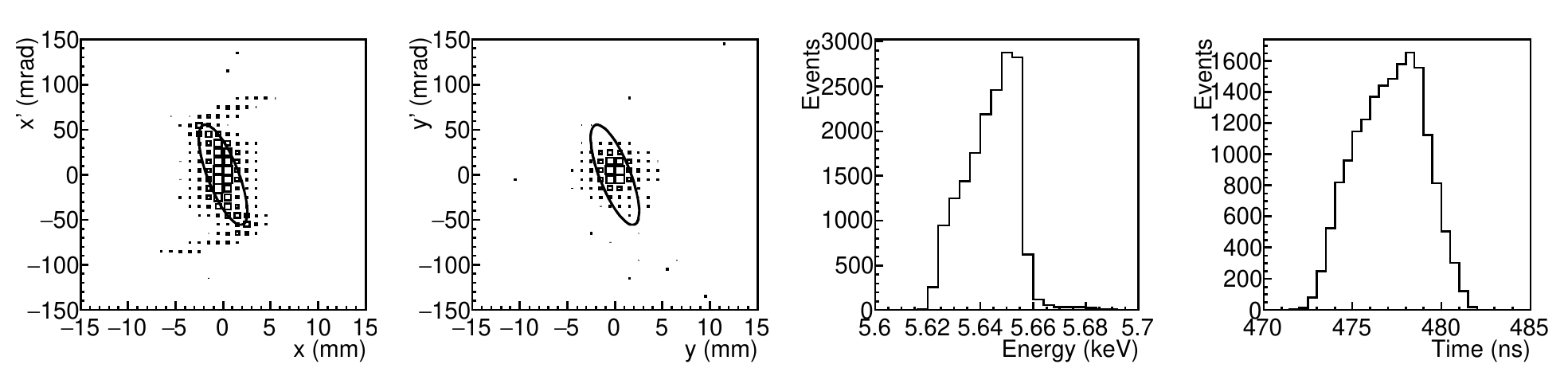}
\caption{Muon beam distribution at the RFQ entrance.
  The ellipses in the transverse distributions represent
  the matched ellipses of 1.0$\pi$~mm~mrad.
}
\label{fig:usm_emitana}
\end{figure}

 As the first acceleration step, thermal muons are accelerated from the ionization region
by a pair of meshed metal plates and an electrostatic lens, a Soa lens~\cite{Soa}.
Figure~\ref{fig:usm_emitana} shows distributions at 
the input of the RFQ linac simulated from distributions of 
the muon source~\cite{otani:simulation_usm_production:ipac2018}.
The ellipses in the $x$-$x^{\prime}$ and $y$-$y^{\prime}$ distributions represent
the matched ellipses of 1.0$\pi$~mm~mrad.
The right panel represents the time structure at the entrance
of the RFQ.
Even though the pulse width of the dissociation laser is 1~ns,
the time width at the RFQ entrance is 10~ns
owing to the spatial distribution at ionization.
Therefore, the beam from the source divides into three bunches during 
the acceleration in the RFQ at the frequency of 324~MHz.
A spare RFQ of the J-PARC linac~\cite{kondo:hptest_rfqII:prstab2013}
will be used as a front-end
structure accelerating the muons to 0.34~MeV~\cite{kondo:simulation_mu_rfq:ipac2015}.
\response{A test of accelerating negative muonium ion is reported in Ref.~\cite{bib:RFQacctest}.}

\newcommand{\Ez}{E_{z}}

The energy of the muon beam is boosted to 4.5~MeV with an IH-DTL.
Different from the Alvarez DTL, the IH-DTL uses the TE11 eigenmode,
and $\pi$-mode acceleration~\cite{bib:IH_blewette}. 
With this mode, the acceleration length is halved compared with
the 2$\pi$-mode acceleration. 
In addition, alternative phase focusing (APF)~\cite{bib:IH_APF1} is adopted.
Since the use of APF eliminates the need for installing quadrupoles
in the drift tubes, a higher shunt impedance per length can be achieved. 
The beam dynamics with such an IH-DTL was studied~\cite{otani:ih_dtl_design:prab2016}.
Sixteen cells are required to accelerate up to 4.5~MeV,
and the total length of the cells is 1.29~m. 
The quality factor $Q_{0}$ is calculated to be $1.03\times 10^{4}$, 
and the power dissipation is 320~kW. 
The effective shunt impedance per unit length is calculated
to be 58~$\mathrm{M\Omega/m}$,
which is competitive with those of other IH structures, taking 
our IH application to a relatively higher velocity region into account. 


Following the IH-DTL, DAW structures
with a frequency of 1,296~MHz are used to accelerate to 40~MeV.
The DAW is one of the coupled-cavity linacs which has large coupling 
between the cells and a high shunt impedance, especially in the middle
$\beta$ section~\cite{andreev:he_proton_linac_structures:linac1972}. 
The cell design was optimized for the velocities of 
$\beta=0.3, 0.4, 0.5,$ and $0.6$ 
by using the SIMPLEX algorithm~\cite{otani:develop_mulinac:ipac2016}.
PARMILA~\cite{parmila} was used to design
the beam dynamics of the DAW section.
The acceleration gradient is determined to be 5.6~MV/m to keep
the maximum electric field less than 0.9 times the Kilpatrick
limit~\cite{bib:kilpatrick}. 
The field strengths of the quadrupole doublets between the modules and
the number of cells in each module are determined
with a condition that the phase advance in one focusing period
is less than 90~degrees. 
The number of cells in a module is set to ten, and  
the phase advance is approximately 83~degrees in the first module,
where the RF defocusing is strongest. 
The total length is 16.3~m with 15~modules.
The estimated power dissipation is 4.5~MW. 

\begin{figure}[t]
\centering
\includegraphics*[width=1.0\textwidth]{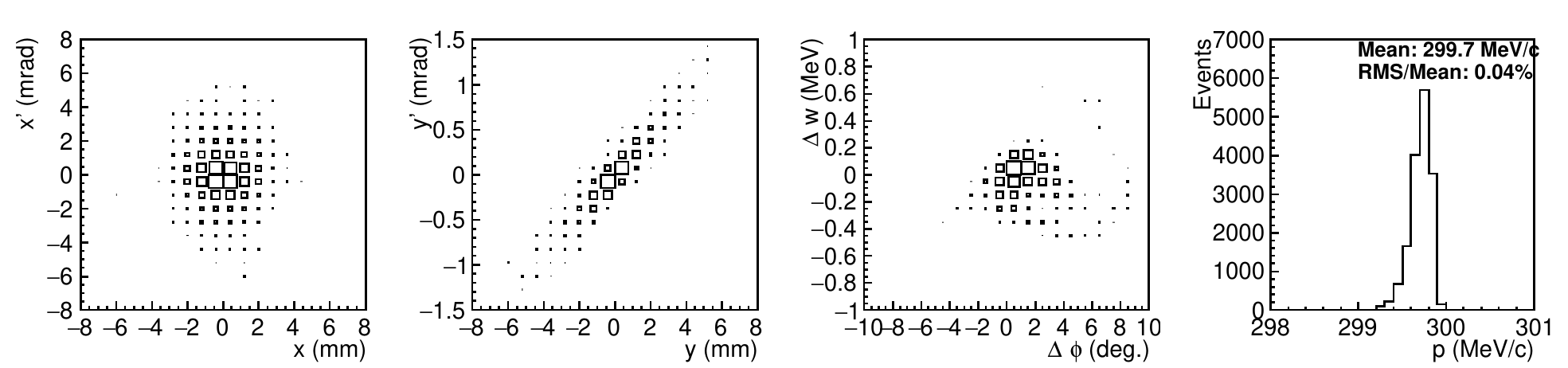}
\caption{Phase-space distributions at the muon linac exit.
  The $\Delta \phi$ and $\Delta$w denote the phase and energy difference
from the synchronized ones.}
\label{fig:dist_dlsout}
\end{figure}

 Finally, the muons are accelerated from 40~MeV to 212~MeV by using a DLS,
which is widely used for electron linacs.
 The advantage of the DLS is its high acceleration gradient;
approximately 20~MV/m.
 An RF frequency of 1,296~MHz is adequate for the wider phase space.
The particular design feature of the DLS for
muon acceleration, which is different from 
the general accelerating structure for an electron accelerator, 
is the variation of the disk spacing corresponding to 
the muon velocity~\cite{kondo:beamdynamics_muon_highbeta:ipac2017}.
 The DLS section consists of four accelerating structures and 
the total length is approximately 10~m.
Figure~\ref{fig:dist_dlsout} shows the phase-space distributions
at the exit of the DLS (muon linac exit) obtained by simulation. 
The estimated momentum spread is 0.04\% (RMS).

The results of the acceleration simulations 
are summarized in Table \ref{tbl:e2e_summary}.
With this design of the muon linac, these simulations show that the transmission efficiency is kept high,
and there is no significant growth of the beam emittance during the acceleration.
The beam pulse width is 10~ns consisting of three microbunches, and the repetition rate is 25~Hz.

\begin{table}[t]
   \centering
   \caption{Summary of the particle simulations through the muon linac}
   \begin{tabular}{|l|r|r|r|r|r|}
     \hline
                                           & Soa & RFQ  & IH   & DAW  & DLS \\
     \hline
     \hline
     Transmission (\%)                     & 87   & 95   & 100  & 100  & 100 \\
     Decay loss (\%)                       & 17   & 19   &  1   &  4   &  1  \\
     $\varepsilon_{n, \, \text{rms}, \, x}$~($\pi$~mm~mrad) & 0.38 & 0.30 & 0.32 & 0.32 & 0.33\\
     $\varepsilon_{n, \, \text{rms}, \, y}$~($\pi$~mm~mrad) & 0.11 & 0.17 & 0.20 & 0.21 & 0.21\\
     \hline
   \end{tabular}
   \label{tbl:e2e_summary}
\end{table}

\section{Beam injection and muon storage magnet}

The muon beam must be injected into the muon storage magnet and the injection system
must have minimum interference to the storage field. For reasons described later,
a new method to inject the muon beam from the top of the magnet is adopted.
After the linac, the muon beam follows a beam transport line to inject the muon 
beam at an incident pitch angle of $-25$~degrees. The beam transport line consists of 
two dipole magnets for bending the beam vertically, three normal quadrupole magnets
to match the vertical momentum dispersion and eight rotated quadrupole magnets
to control the phase space to match the acceptance of injection into the magnet.

\begin{figure}[t]
 \centerline{\includegraphics[width=12cm]{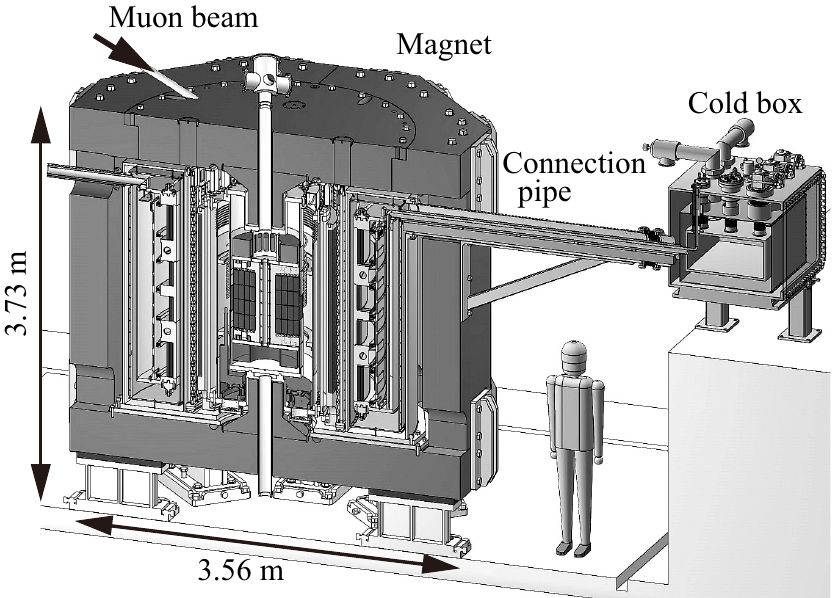}}
 \caption{Overview of the muon storage magnet.}
\label{fig:magdesign}
\end{figure}

A $3$~T MRI-type superconducting solenoid magnet will be used to complete the injection and store the muon beam.
Figure~\ref{fig:magdesign} shows an overview of the muon storage magnet~\cite{sasaki:2016}.
The muons are stored in a 3~T magnetic field with a cyclotron radius of $333$~mm.
This cyclotron radius is about a factor of 20 smaller than that for the BNL/Fermilab experiments.
We take advantage of the advance in MRI magnet technology to fabricate such a small storage magnet with a highly uniform magnetic field in the muon storage region.
As summarized in Table~\ref{tab:function}, the magnet system has four functions: (1) provide a highly uniform storage field, (2) provide the injection field, (3) provide the kicker field to store the muons, and (4) provide weak focusing for storage.

\begin{figure}[t]
 \centerline{\includegraphics[width=10cm]{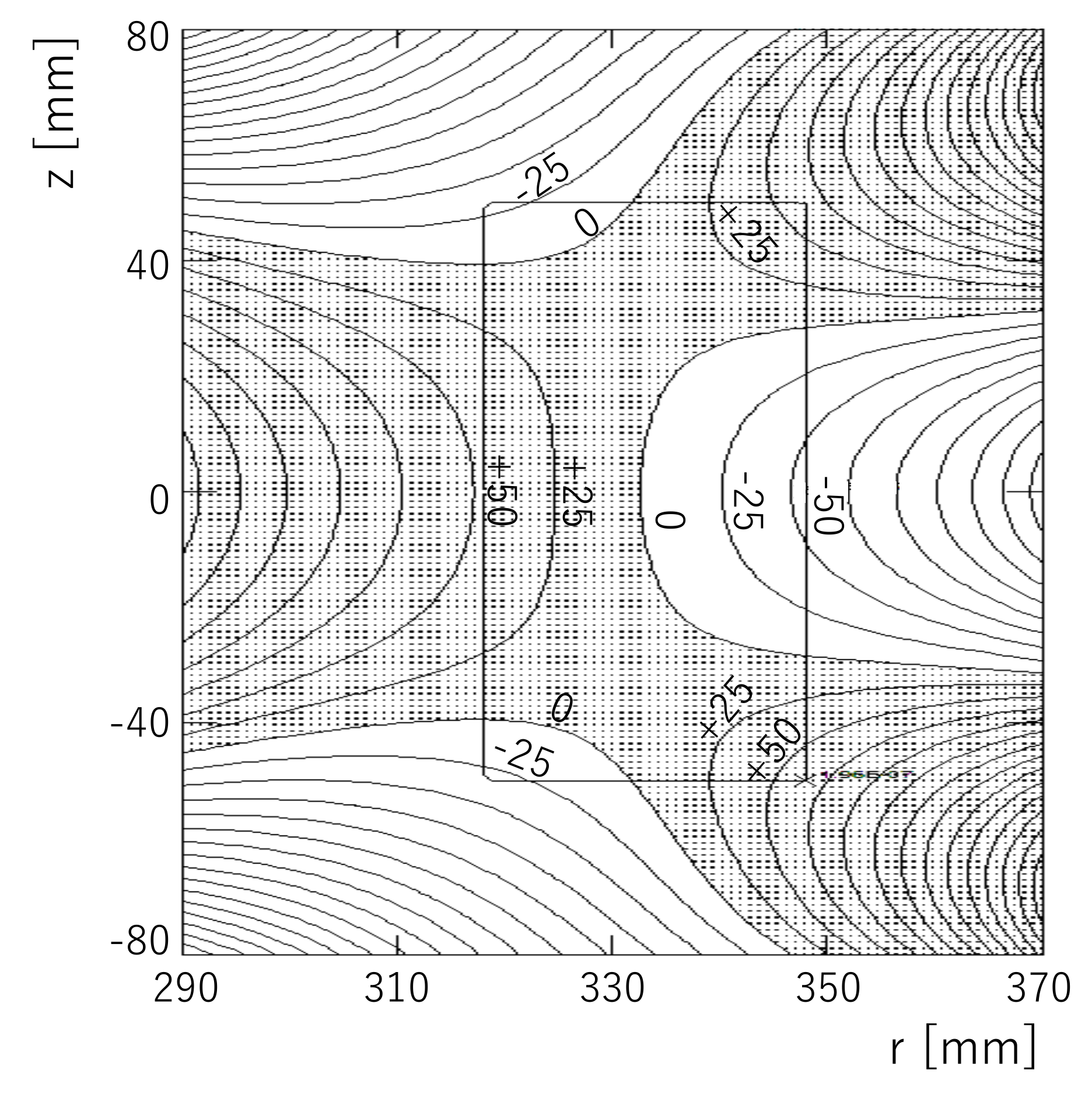}}
 \caption{Designed distribution of the main magnetic field relative to the reference field ($B_0=3$~T) averaged over the storage ring. 
 	In the dotted area, $B$ is larger than $B_0$.
          Contour lines of residual magnetic field are at $B_0$, and $B_0\pm$ every $25$~ppb ($0.075$~$\mu$T).  
          The inset rectangle is the region of the stored muon orbit.
          The numbers in the figure are the residual magnetic field strengths in ppb.
           See more details in~Ref.~\cite{NIM-Abe}.}
\label{fig:abe-san}
\end{figure}

The main feature of the magnet is a highly homogeneous magnetic field of $3$~T (main field) 
in the central region of the magnet, the storage region, where the muon beam is stored until its decay. 
The homogeneity of magnetic field in the storage region is directly related to the sensitivity of the $a_{\mu}$ measurement.
The integrated main magnetic field uniformity along the beam orbit in the storage region has to
be carefully controlled with a precision of $100$~ppb peak-to-peak.
Figure~\ref{fig:abe-san} depicts the estimated relative field distribution in the $r$-$z$ plane around the storage region averaged over the storage ring,
\response{where the $z$-axis is the center axis of the magnet along the direction of magnetic field and $r$ is the distance from the $z$-axis in a plane perpendicular to the axis.}
Averaged over the muon orbit along azimuthally, the variation is estimated to be $\pm 50$~ppb. 
The average field variation along the muon orbit for the BNL~(E$821$) magnet was as large as $\pm$500 ppb~\cite{Bennett:2006fi}.

\begin{figure}[t]
   \centering
   \includegraphics*[width=10cm]{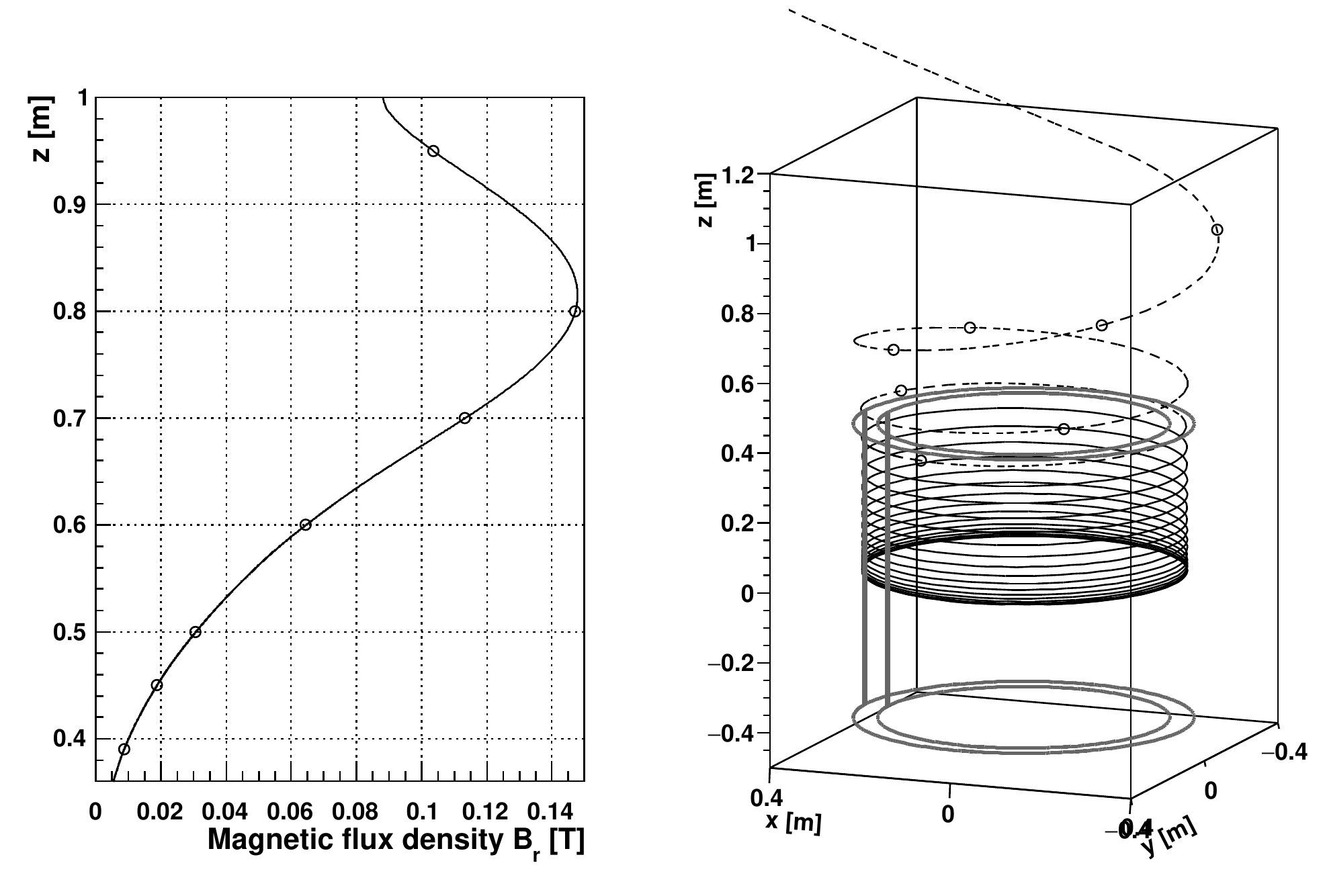}
   \caption{Outline of the three-dimensional injection scheme.  
          The muon beam enters the solenoid obliquely from above into the injection region~(the solenoid fringe field). 
            Left:~Radial component of the fringe field in the injection volume.
            Right:~Three-dimensional view of the beam trajectories from the injection through the storage.   
            A dotted line shows a design trajectory the injection region.
             Open circles along the trajectory indicate in the corresponding positions in the left plot.  
            A solid line shows a design trajectory in the kicker region.
            Two pairs of one-turn coils for the kicker, which store the beam, are also shown.
   }
   \label{BR-INJ}
\end{figure}

The second function is to transport the muon beam from the outside of the storage magnet to the storage region. 
This transportation region is named the injection region.
Due to the limited space of the storage magnet, the muon beam is not injected by the method used in the previous experiments of horizontal injection using an inflector magnet. 
Instead, a new $3$-D spiral injection scheme~\cite{NIM-3D}, as displayed in Fig.~\ref{BR-INJ},
is developed for this purpose.

A solenoid magnetic field shape is suitable for this new injection scheme. 
In the injection region, the radial component, $B_{r}$, of the magnetic field has to be carefully controlled from the top end of the magnet to the storage region for smooth injection. 
The left panel of Fig.~\ref{BR-INJ} depicts the radial component of the fringe field along the beam in the injection volume.
A three-dimensional view of beam trajectories from the injection region to the storage region is also shown in the right panel. 
The muon beam is injected with pitch angle of $440$~mrad. 

Open circles along the beam indicate points that correspond to the radial field values on the left panel.  
The beam momentum is deflected by 
$B_{r}$ as it reaches the mid-plane of the solenoid magnet. Within the first three turns, the pitch angle becomes $40$~mrad.  
We design the fringe field to control beam vertical motion.
And at the same time, this fringe field requires  
appropriate vertical-horizontal coupling (so-called $X$-$Y$~coupling in the beam coordinate) to control vertical divergence, 
because of an axial symmetric shape in the fringe field. 
The $X$-$Y$~coupling of the beam phase space, controlled by the magnets located just upstream of the solenoid, will be carefully tuned to 
minimize the vertical beam size in the storage region. 

\begin{figure}[t]
   \centering
   \includegraphics*[width=10cm]{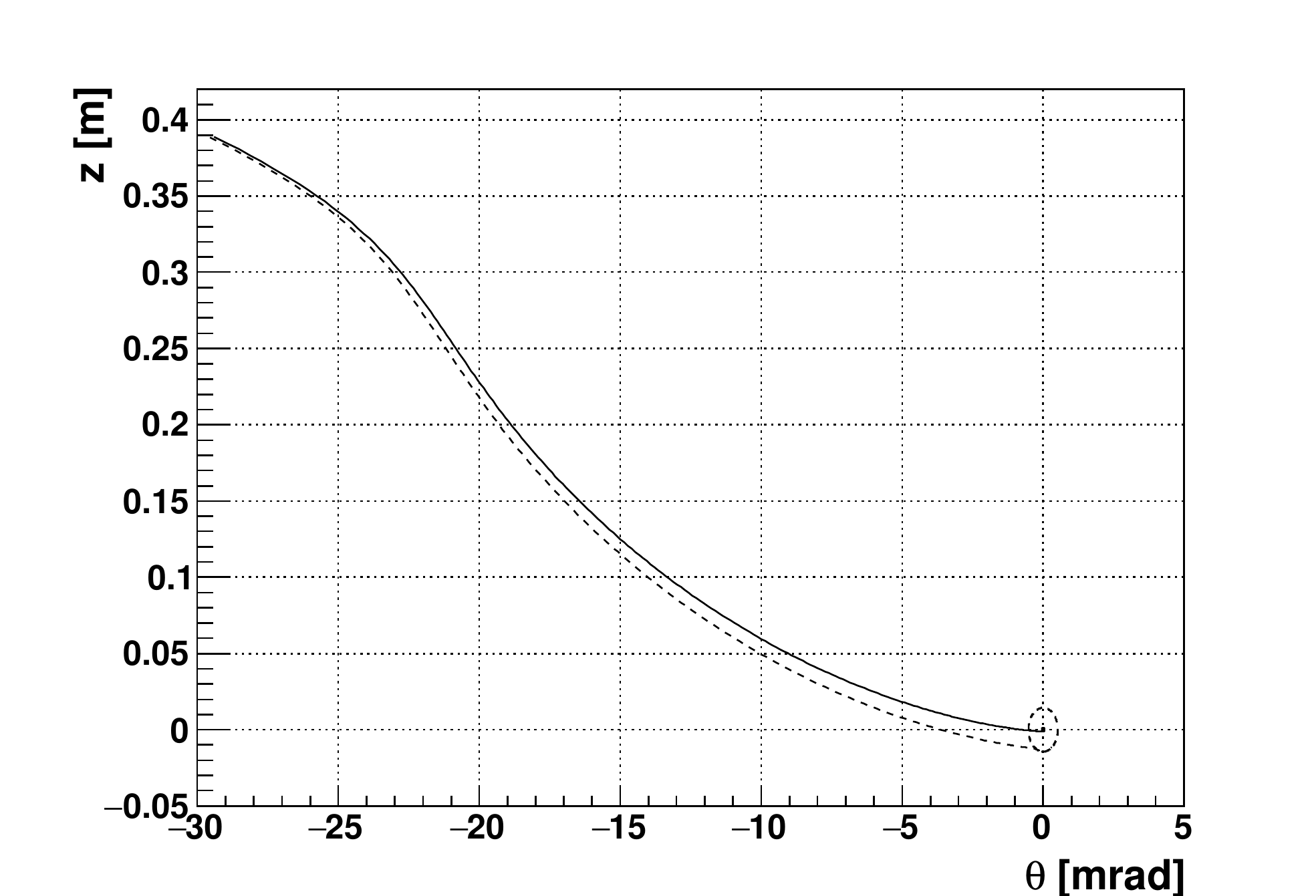}
   \caption{
      Vertical beam motion during and after the kick for sample trajectories.
      The vertical axis is the vertical position and the horizontal axis is the pitch angle.   
      The solid line is a design trajectory for the center of the beam.
      In the case that the muon does not stop on the mid-plane ($z=0$~m) at the end of the kick, 
      the muon will stay within the closed ellipse due to the weak focusing field,
      shown as a dotted line.   
    }
   \label{kick}
\end{figure}

The third function of the magnet system is to provide a vertical kick, which will guide the beam inside the storage region.
Two pairs of one-turn coils, the kicker coils positioned at heights of $\pm0.4$~m, 
generate a pulsed radial field $B_{\text{kick}}$ to apply a vertical kick to the muon beam motion. 
Figure~\ref{kick} shows the vertical beam motion from the start of the kick to the end, as well as the beam motion in the storage region.

The weak focusing field is the fourth function of the magnet system.
In order to keep the beam inside the storage region within a stable orbit,
a weak focusing magnetic field~\cite{NIM-3D,NIM-Abe} will be used.
The equations of the weak focusing magnetic field are 
\begin{eqnarray}
 B_r &=& -n\frac{B_{0z}}{R}z, \\
 B_z &=& B_{0z}-n\frac{B_{0z}}{R}(r-R)+ n\frac{B_{0z}}{2R^2}z^2,  
\label{eq:weak}
\end{eqnarray}
where, $B_{0z}$ (3~T) is the field strength in the $z$ direction at the center of the storage region,
$R$ (333~mm) is the average radius of the stored beam, $n$ is the field index.


\begin{table}[t]
\centering
    \caption{ Functions and specifications of the magnet system }
  \begin{tabular}{|l|l|l|}\hline 
    Functions       & Location & Specifications   \\ \hline \hline
    Main field      & $r = 333 \pm 15$~mm,   & Axial field ($B_{0z}$) = $3$~T       \\
                    & $z = \pm 50$~mm        & Local uniformity  $<$ 1,000 ppb      \\
                    &                        & Integrated uniformity along the orbit \\ 
                    &                        &    less than 100~ppb (peak-to-peak) \\ \hline
    Injection field & $0.4 < z < 1.1$~m      & Radial field with $B_r \times B_z > 0$           \\  \hline
    Kicker field    &  $|z|<0.4$~m           & Radial pulsed field created by  \\
                    &                        & two pairs of round-type kicker coils.\\ \hline
    Storage field   & $r = 333 \pm 15$~mm,   & Weak magnetic focusing, \\
                    & $z = \pm 50$~mm        & n-index $\sim (1.5 \pm 0.5) \times 10^{-4}$ \\ \hline 
  \end{tabular}
     \label{tab:function}
\end{table}

The solenoid will be composed of five main coils wound with NbTi cable
and the inner radius will be 0.8~m in the present design.
An iron yoke is used to suppress magnetic flux leakage.
The magnet has pole tips at both ends of the solenoid coil to form the magnetic flux, with an entrance hole for injection.

The main coils will be operated in persistent current mode (PC mode) with a superconducting switch.
The time constant of current decay during the nominal operation is generally expected to be less than 10 ppb/hour.

The weak focusing magnetic field is generated by dedicated coils, the weak focus coils, consisting
 of eight ring coils wound with NbTi cable that are aligned in the axial direction of the magnet.
All ring coils are connected in series electrically and driven by a single power supply.

The magnetic field is shimmed by passive and active shimming systems.
The former uses iron pieces,
which are attached on support cylinders installed inside the magnet bore through holes in the iron poles in air.
The magnetic field distribution is adjusted by changing the alignment pattern of iron pieces.
Active shimming is done using superconducting shim coils wound with NbTi cable.
They are mainly used to compensate the error field changing with time in the storage region,
and the residual error (expected to be small) after the magnetic field shimming by iron pieces.
The shim coils consist of several saddle coils which have a four-fold symmetry.
Each coil is connected to an independent power supply to control each current.

The main coils, the weak focus coils, and the shim coils are immersed in liquid helium to ensure good temperature stability.
The helium is recondensed by cryocoolers for long-term, stable and cost-effective operation.
Four cryocoolers and a heat exchanger for the helium recondensation will be installed in a cold box,
 placed apart from the magnet cryostat.
A connection pipe between the cold box and the magnet cryostat has a bellows connection,
which is a soft connection in terms of mechanical structure, so that
the vibration of cryocoolers will not be directly transferred to the magnet.

The magnetic field in the storage region is measured by a nuclear magnetic resonance (NMR) probe.
A continuous-wave NMR (CW-NMR) magnetometer will be used in \response{our} experiment. 
The resonant absorption signal of protons $\omega_{p}$ in water samples is observed
by using a fixed frequency source and a small sweeping magnetic field.
The NMR probe will have a size of about 5--10~mm in diameter.
Several NMR probes will be mounted on the three-axis moving stage in radial, azimuthal and vertical directions 
to scan the storage region for the magnetic field mapping.
The mapping probes are evacuated from the inside to outside of the storage region during the muon beam storage.
The stages are driven by ultrasonic motors, which can work in the strong magnetic field.
The ultrasonic motors have encoders so that the position of the NMR probe is controlled
with a precision of below 0.1~mm.

In addition to the mapping probes, several other NMR probes will be installed below the storage region to measure time variation of the magnetic field strength, the fixed probes. 
The magnetic field strength will slightly and steadily decay in the PC mode, as described above, 
and it will also slightly fluctuate due to temperature variations.
In order to compensate such small fluctuations of the magnetic field, and to know the best timing for the magnetic field restoration,
we monitor the time variation of the magnetic field continuously at several appropriate field positions.
The fixed probes do not monitor the deformation of the magnetic field distribution in the storage region but its time variations. 
A correlation between the magnetic field deformation and the field strength will be measured during the commissioning period for the detailed compensation of $\omega_{p}$. 

\section{Positron Detector}\label{sec:detector}

The positron detector is installed inside the storage magnet
and measures positron tracks from decay of the stored muon beam.
\response{The muon storage region is kept in high vacuum not to cause beam emittance growth while 
the detector region is separated from the storage region by a polyimide film and is kept in 
medium vacuum.}
A muon with momentum of 300~MeV/$c$ circulates with a radius
of 333~mm and decays to a positron, a neutrino and an antineutrino
with a dilated  lifetime of 6.6~$\mu$s. The cyclotron period is 7.4~ns.
Since the anomalous precession period is 2.1~$\mu$s, muons circulate
the ring about 300 times on average during one revolution of muon spin.
The goals of the detector are to measure $\omega_{a}$
and the up-down asymmetry of positron direction due to EDM.

Due to non-conservation of parity in the weak decay of muons, the average positron energy
is higher when positrons are emitted closer to the muon spin~\cite{Michel}. 
By measuring high energy positrons
selectively, positrons emitted forward can be selected and
the time variation of muon spin with respect to the muon momentum direction
can be measured. The sensitivity becomes maximum when positrons
with momentum above 200~MeV/$c$ are counted.
The maximum momentum of decay positrons is 309~MeV/$c$ while the momentum in the range
from 200~MeV/$c$ to 275~MeV/$c$ will be used for the analysis.

\begin{figure}[tbp]
  \begin{center}
    \includegraphics[width=0.4\textwidth]{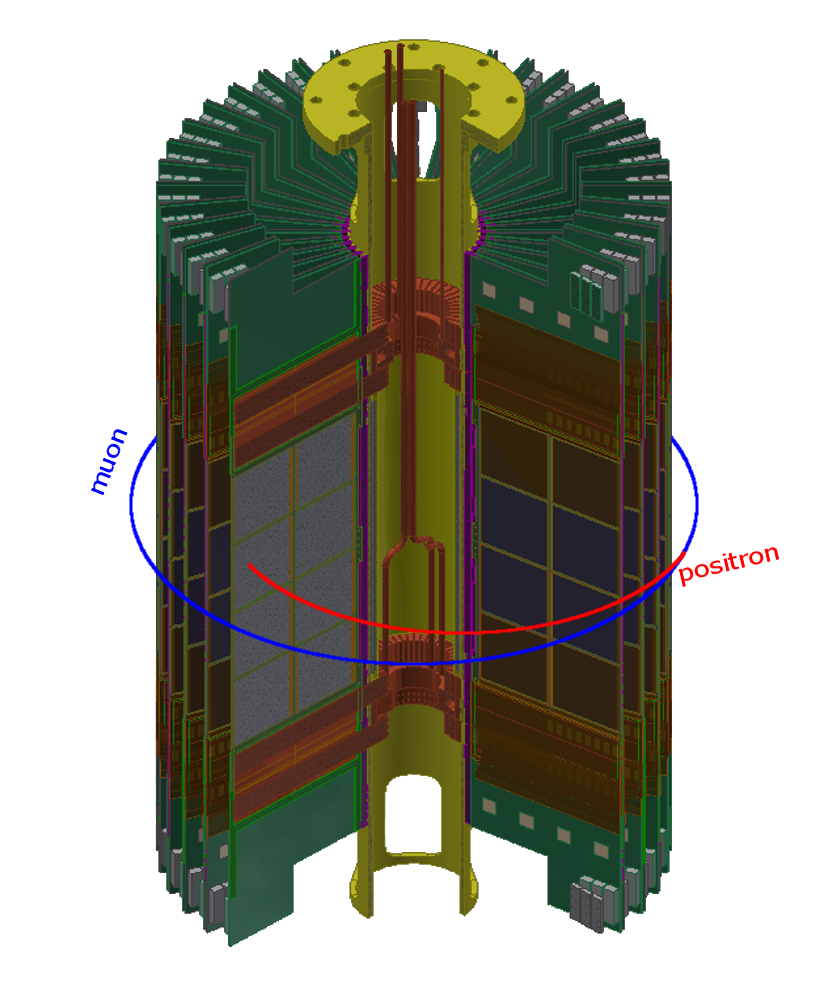}
    \includegraphics[width=0.4\textwidth]{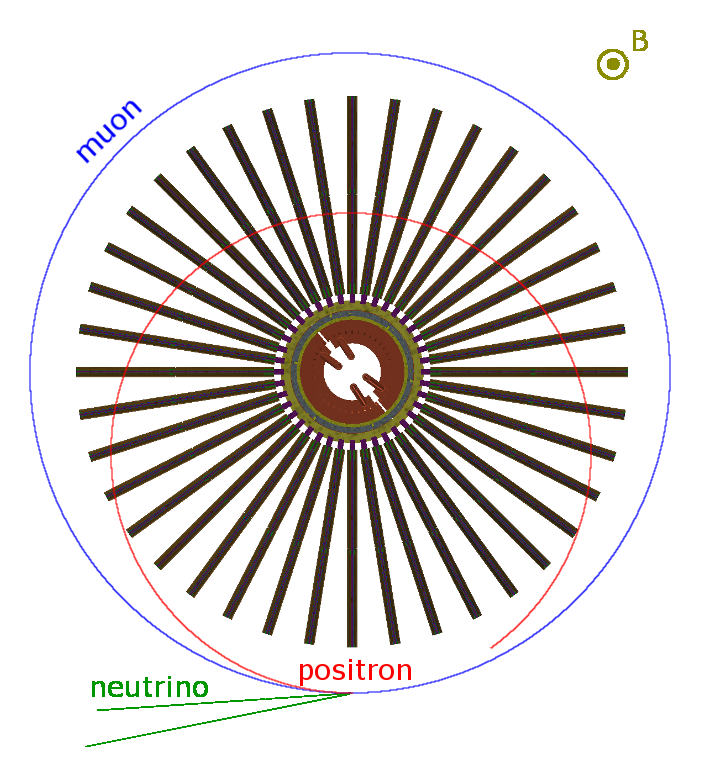}
    \caption{Perspective view (left) and top view (right) of the positron detector.}
    \label{fig:Detector_overview}
  \end{center}
\end{figure}

Positrons emitted within the 3~T magnetic field move in a spiral orbit.
This trajectory is detected by radially arranged silicon strip sensors.
Geometrical coverage of the detector is 90--290~mm in radial direction and within $\pm$200~mm
in height. The layout of the detector is shown in 
Fig.~\ref{fig:Detector_overview}. 

The muon beam time structure following acceleration to 300~MeV/$c$
is a pulse of 10~ns width consisting of three microbunches, with
a repetition rate of 25~Hz. This is the time structure of the
fill of the muon storage ring. The number of muons per fill is
about $10^4$. The measurement will be performed in an interval
following the fill of 33~$\mu$s, which is five times larger than the
time-dilated muon lifetime.
The rate of positrons
changes by a factor of 160 from the beginning to the end of the measurement.
Thus, the detector is required to be stable against the change of positron rate;
otherwise, the measured $\omega_a$ would be biased.

The detector consists of 40 radial modules called vanes.
Each vane consists of 16 sensors, half of which measure the
radial coordinate and half the axial coordinate of ionization.
Sensors are made by single-sided p-on-n silicon technology~\cite{sensor}.
The active area of a sensor is 97.28~mm $\times$ 97.28~mm with
a thickness of 0.32~mm.
A sensor has two blocks of 512 strips with a pitch of 190~$\mu$m. 
Therefore a vane has 16,384 strips, with 655k total strips for the detector.


The data from the silicon strip sensors are read out by front-end
boards with Application Specific Integrated Circuit (ASIC) on the detector with a 5~ns time stamp, followed by readout boards with 
Versa Module Eurocard (VME) interface,
then collected by the PC farm through a Gigabit Ethernet switch.
The data acquisition system is based on DAQ-Middleware~\cite{DAQ-Middleware}.
The estimated rate of data from the whole detector is 360~MB/s (or 14.4~MB/fill).

One readout ASIC has 128 channels for analog and digital blocks.
The dynamic range of input charge is required to be greater than
four minimum ionizing particles (MIP) equivalent with linearity.
Equivalent noise charge is required to be less than 
1600~e$^{-}$ with input capacitance of 30~pF, which
corresponds to signal-to-noise ratio greater than 15
for a 1~MIP signal.
One of major systematic uncertainties on $\omega_{a}$ is hit timing shift due to pile-up hits. 
If several charged particles pass through the same sensor strip within the pulse width, 
signal pulse shape is distorted and the detected timing shifts. 
Since the pile-up rate changes as a function of time, 
this timing shift causes a systematic shift of $\omega_{a}$ measurement.
To constrain this effect, the peaking time at 1~MIP charge is required to be less than
50~ns and the time-walk between 0.5~MIP and 3~MIP is required to be less than 5~ns.

The system clock is provided by the Global Positioning System (GPS)-synchronized Rb frequency
standard~\cite{freqtime}, and it is distributed with real time control signals
to the readout boards and the front-end board through the timing
control/monitor board. Long term stability of the system clock
frequency is confirmed better than $10^{-11}$.





The stringent requirement on detector alignment comes
from the EDM measurement~\cite{EDM_requirement}. Alignment accuracies of vanes
with respect to the magnetic field direction
are required to be better than 10~$\mu$rad for skew, i.e., the
angle around an axis normal to the vane. 
In order to ensure the required accuracies, alignment changes for the
vanes are detected and monitored during operation using an absolute
distance interferometer system~\cite{IWAA}.

\begin{figure}[htbp]
  \begin{center}
    \includegraphics[width=0.7\textwidth, angle=0]{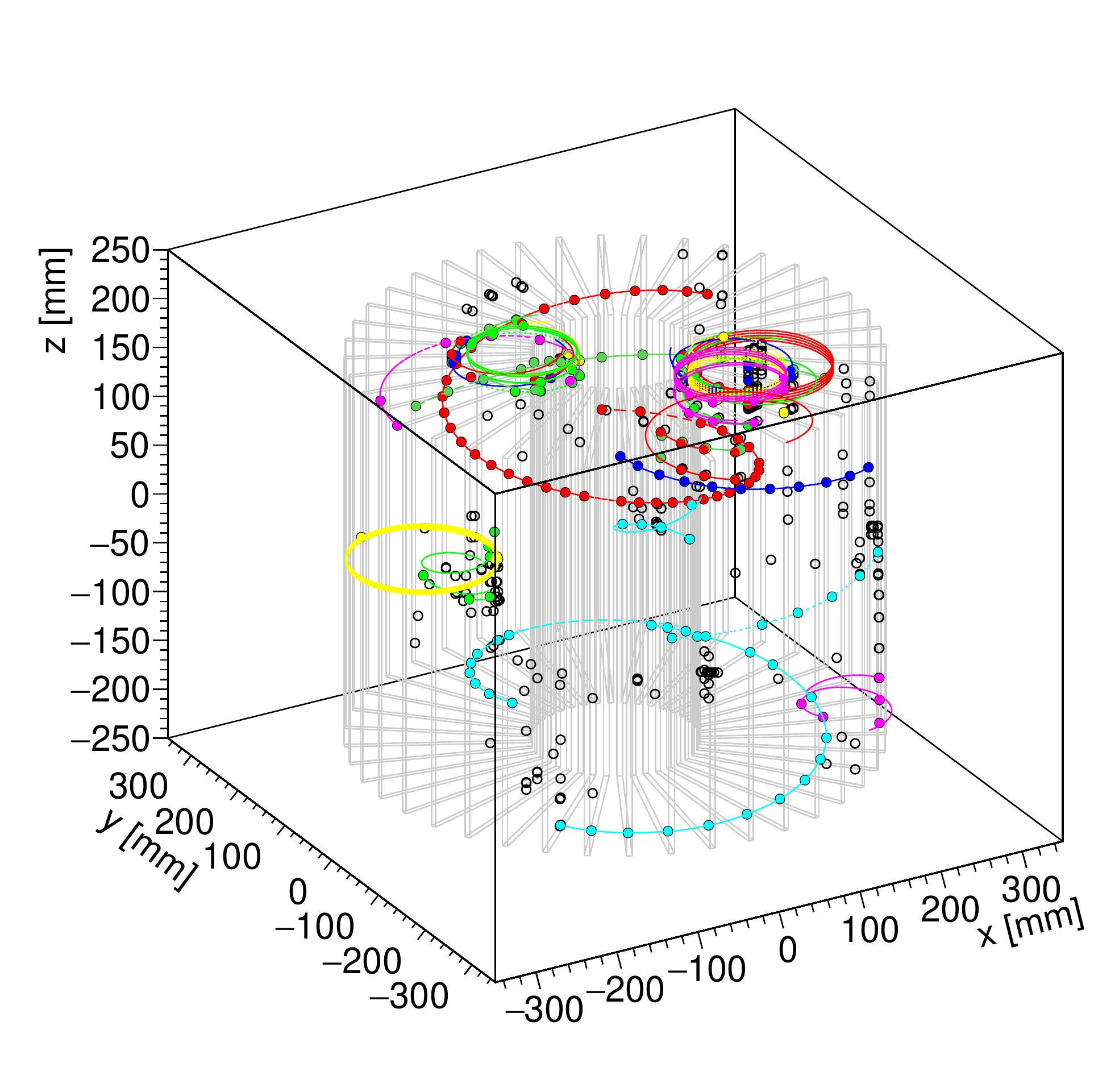}
    \includegraphics[width=0.45\textwidth, angle=0]{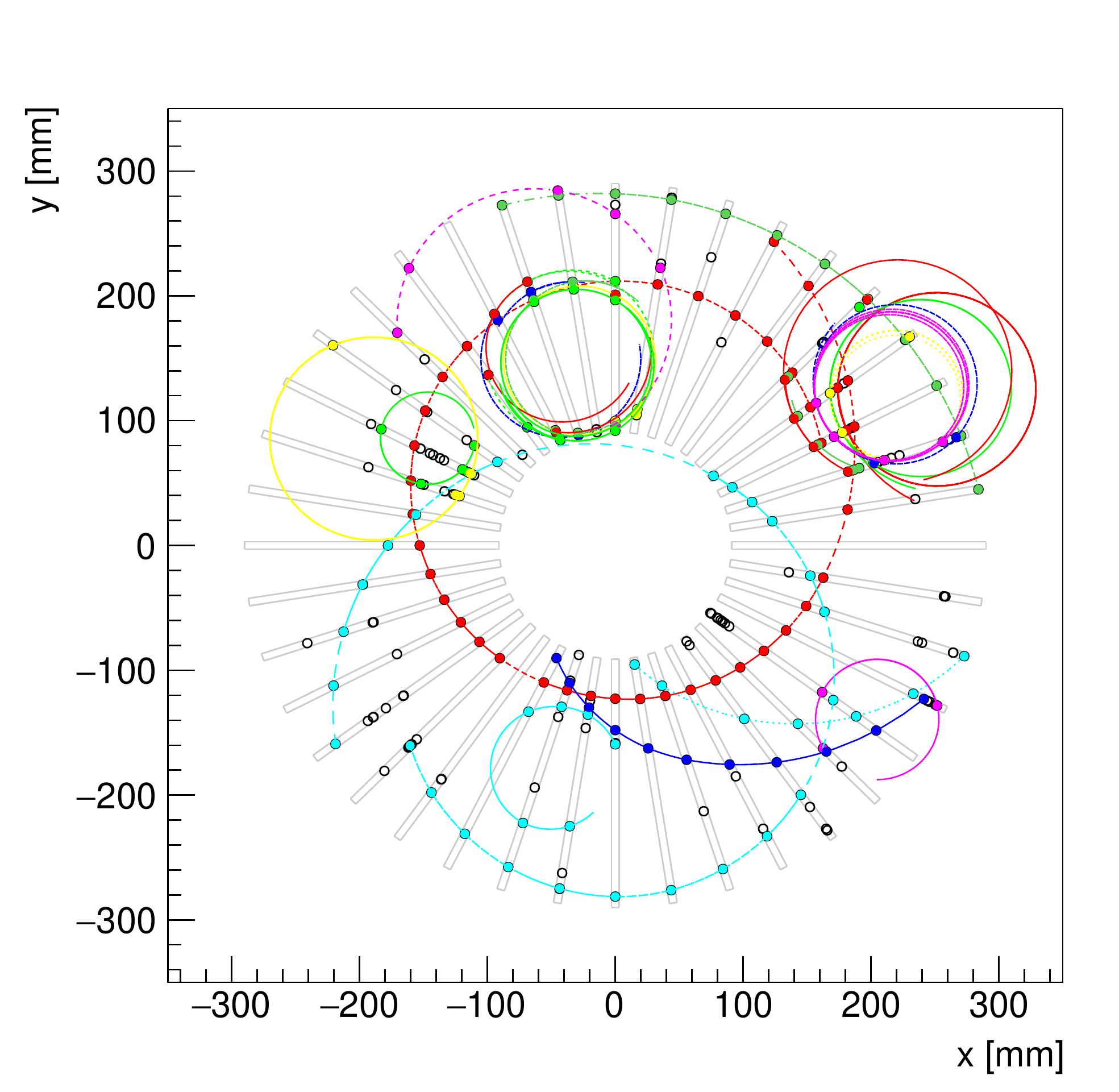}
    \includegraphics[width=0.45\textwidth, angle=0]{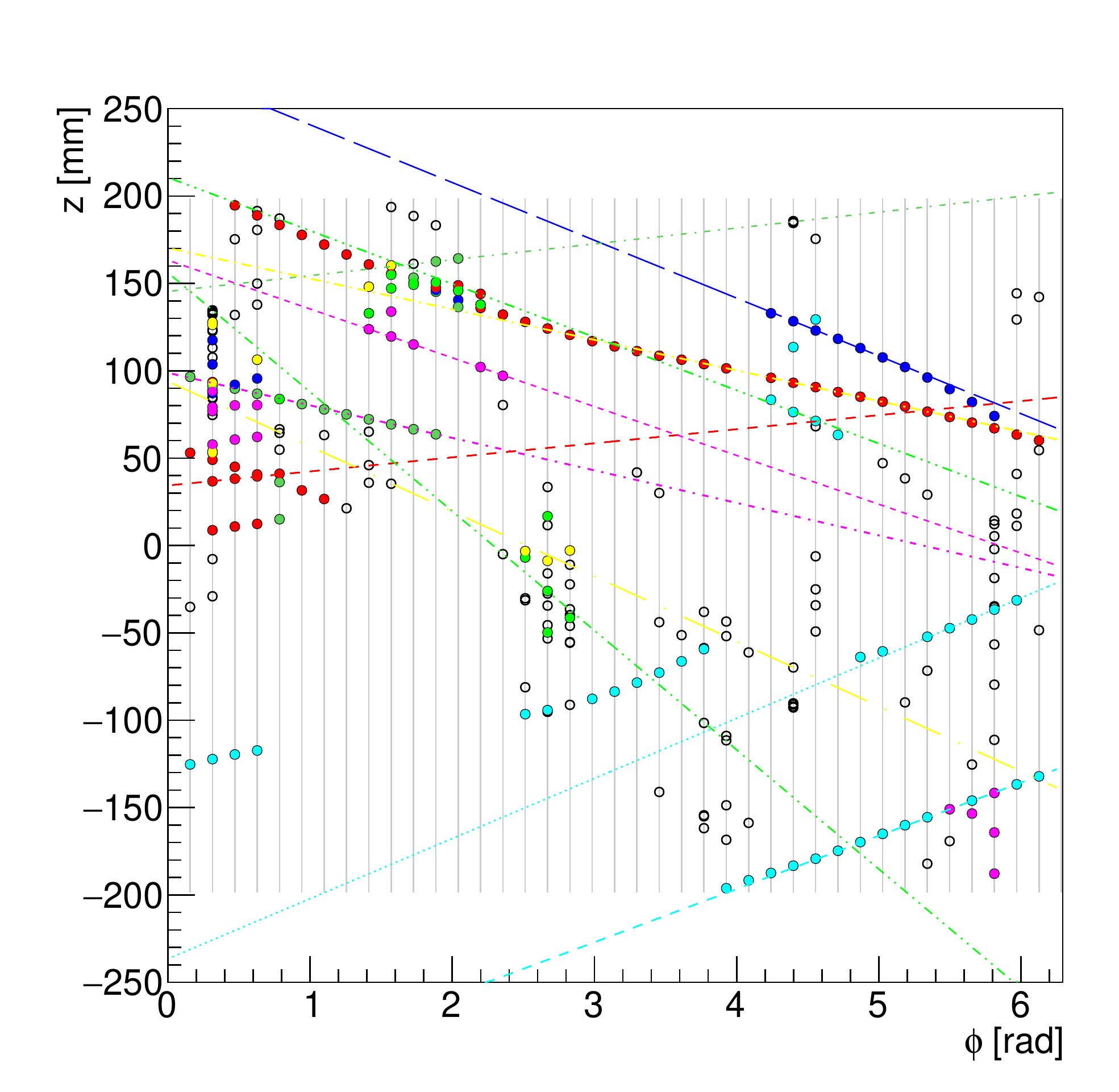}
    \caption{
	All reconstructed hits from 25 muon decays obtained from simulation 
	projected onto the horizontal ($x$-$y$) plane (bottom left) 
	and in the $\phi$-$z$ plane (bottom right), and perspective view in three-dimensional space (top) are shown. 
	There are two positron tracks in the energy range of 
	$200~{\rm MeV} < E< 275~{\rm MeV}$. 
	Track candidate hits are shown by colored dots and the other hits are shown by white dots. 
	Reconstructed track orbits are shown by colored curves (top and bottom left) and straight lines 
	for track finding are shown (bottom right).
    }
    \label{fig:track_finding}
  \end{center}
\end{figure}

At the beginning of the interval after the fill, about 30 positrons are produced from muon decay in 5~ns, which is one time window of the data taking.
The maximum hit rate per silicon sensor strip is $7 \times 10^{-3}$ per time stamp.
To find positron tracks in such a condition,
a positron track candidate is identified from hits in the detector
using the property that high momentum positron tracks leave
nearly straight lines in the $\phi$-$z$ plane, 
where $\phi$ is the angle around the $z$-axis.
Figure~\ref{fig:track_finding} shows event displays and reconstructed tracks obtained from simulation. 
In the $\phi$-$z$ plane (bottom right), straight lines used as seeds for track finding are shown.
A Hough transformation~\cite{HoughTransform,UseHough} is used
to find straight lines in the plane and hits on a straight line are used as the seed. 
A track momentum is obtained
by track fitting with a Kalman filter~\cite{KalmanFilter}.
With this algorithm, a track reconstruction efficiency
greater than 90\% is achieved in the positron energy range of
$200~{\rm MeV} <E< 275~{\rm MeV}$ even at the highest positron rate.

The muon decay position is
determined by the closest point of approach between the reconstructed positron
trajectory and the muon beam orbit.
The muon decay time at the decay position is measured by extrapolating the time of hits in reconstructed
positron tracks. 
One way to estimate decay time is to use the average time of reconstructed 
track hits.
Another approach is to use the transition timing of hits with the 5~ns time stamp
\response{when detector hits are distributed with a width larger than one time stamp.}
The latter method has better timing resolution than the former but
it is applicable only when the transition occurs within a track.
Two definitions of decay time can be cross-checked with each other.

\section{Estimation of the number of reconstructed positrons}\label{sec:Intensity} 
Efficiencies of steps from the surface muon production to the detection of positrons are studied 
by a chain of simulations.
Table~\ref{tab:beamintensity} shows the breakdown of the efficiencies.
The simulations include surface muon production,
thermal muon production, reacceleration, injection to the muon storage
magnet, muon beam dynamics in storage, and finally the detection of the positron.
The simulation of surface muon production~\cite{otani:simulation_usm_production:ipac2018}
and thermal muon production is optimized by the experimental data
on surface muon yield at the existing beamline and measurements of the muonium space-time distribution~\cite{beer:2014},
respectively.
The total efficiency is $1.3 \times 10^{-5}$ per initial muon at production.
At a proton beam power of 1~MW, the expected number of positrons is $5.7 \times 10^{11}$ for $2.2 \times 10^7$ seconds data taking.

\begin{table}[t]
  \centering
  \caption{Breakdown of estimated efficiency}
    \begin{tabular}{|  p{5.2cm}   r||  p{4.7cm}   r|}
    \hline
    Subsystem & Efficiency  & Subsystem & Efficiency  \\
    \hline
    \hline
    
     H-line acceptance and transmission       & 0.16   &    DAW decay                            & 0.96 \\ 
    Mu emission                          & 0.0034 &   DLS transmission                     & 1.00 \\  
    Laser ionization                     & 0.73   &   DLS decay                            & 0.99 \\  
    Metal mesh                           & 0.78   &  Injection transmission               & 0.85 \\    
    Initial acceleration transmission and decay
                                         & 0.72   &   Injection decay                      & 0.99 \\  
    RFQ transmission                     & 0.95   &  Kicker decay                         & 0.93 \\    
    RFQ decay                            & 0.81   &   $e^+$ energy window & 0.12  \\  
    IH transmission                      & 0.99   &     Detector acceptance of $e^{+}$       & 1.00 \\
    IH decay                             & 0.99   &     Reconstruction efficiency            & 0.90 \\
DAW transmission                     & 1.00  & & \\

    \hline
    \end{tabular}
  \label{tab:beamintensity}
\end{table}

\section{Extraction of $a_{\mu}$ and EDM}\label{sec:Sensitivities} 

\begin{figure}[t]
  \centering
    \begin{tabular}{c}

      \begin{minipage}{0.5\hsize}
        \includegraphics[width=0.7\linewidth, angle=0, bb=0 0 408 367]{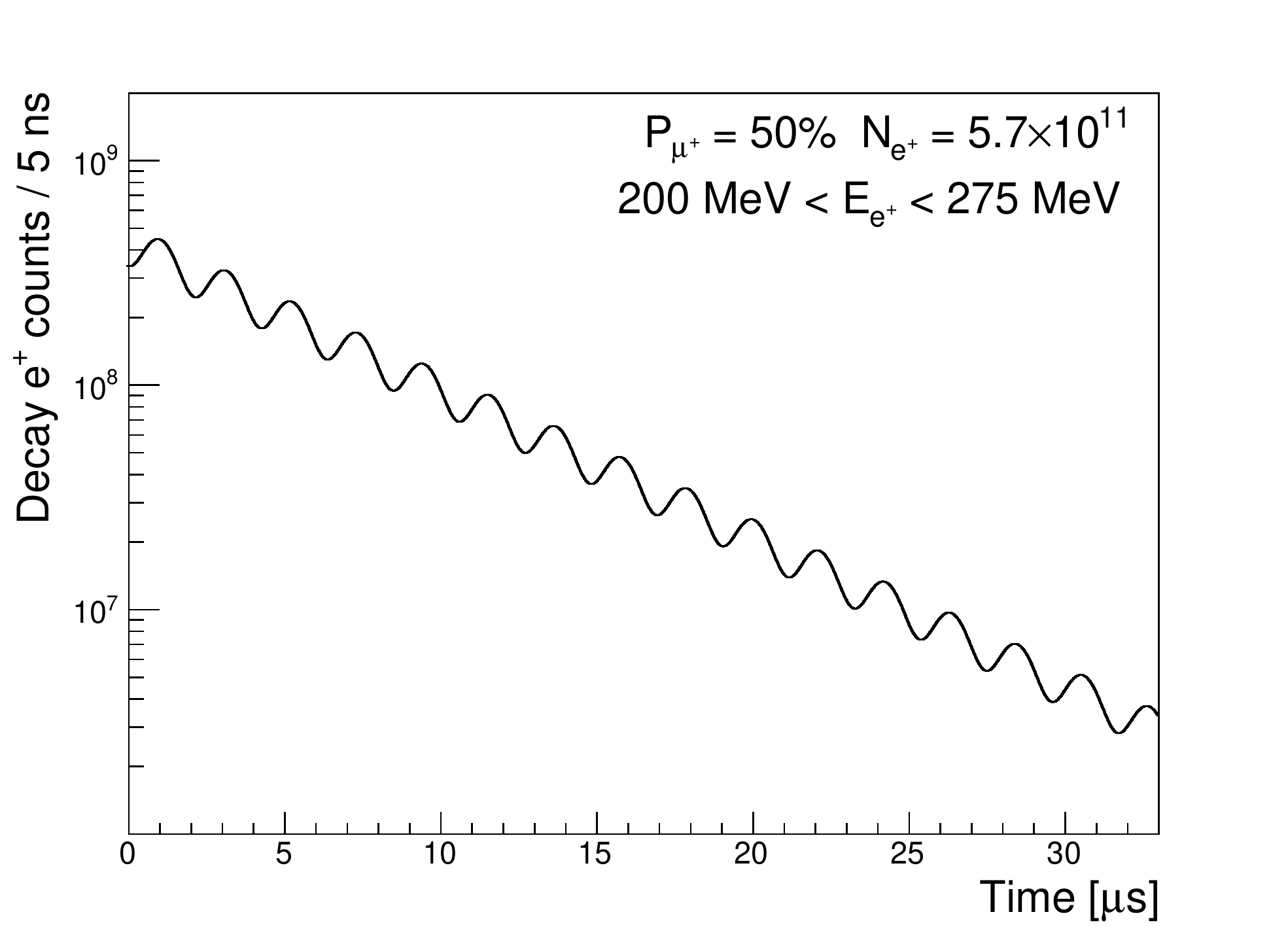}
      \end{minipage}

      \begin{minipage}{0.5\hsize}
        \includegraphics[width=0.7\linewidth, angle=0, bb=0 0 408 367]{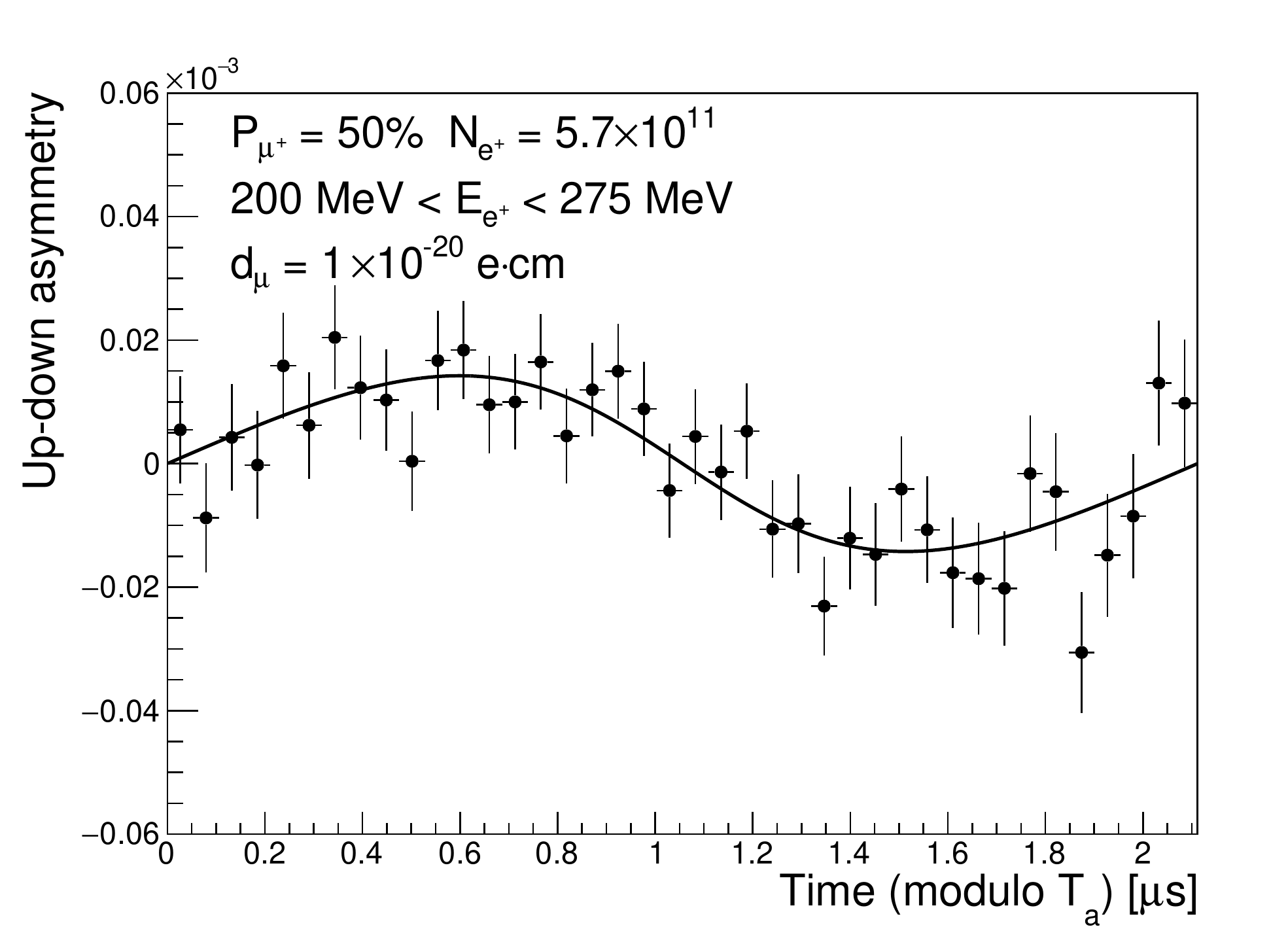}
      \end{minipage}
    \end{tabular}

    \caption{Simulated time distribution of reconstructed positrons (left) 
      and the up-down asymmery as a function of time modulo of the $g-2$ period (right).
      The solid curve is the fit to simulated data.}
    \label{fig:Wiggle}
\end{figure}

The values of $\omega_a$ and $\eta$ are obtained from the muon decay time distribution.
The muon decay time is reconstructed from the positron track as described in Sec.~\ref{sec:detector}.
A simulated time spectrum for detected positrons in the energy range 
between 200~MeV and 275~MeV is shown in Fig.~\ref{fig:Wiggle}~\rev{(left)}.
The anomalous precession frequency $\omega_a$ is extracted by fitting to the data. 
Alternatively, one can make a ratio of 
data taken with opposite initial spin orientations.
This will be useful to study early-to-late changes in the detector performance.

The value of $\omega_p$, from which we determine the average magnetic field seen by the muons in the storage ring, is 
measured by independent measurements of the magnetic field map in the storage ring provided from the proton
NMR data and the muon beam distribution deduced from tracing back the positron track to the muon beam.
A blind analysis will be done as was done in the previous BNL experiment, separating the results for 
magnetic field and spin precession until all systematic uncertainties are finalized.

After the $\omega_a$ and $\omega_p$ are extracted from the experimental
data, $a_\mu$ is obtained from Eq.~(\ref{eq:amu}).
Table~\ref{tab:sensitivity} summarizes statistics and uncertainties for $2.2 \times 10^7$ seconds of data taking.
The estimated statistical uncertainty on $\omega_a$ is 450~ppb, 
while the statistical uncertainty on $\omega_p$ will be negligibly small. 
Thus, the statistical uncertainty of $a_{\mu}$ would be 450~ppb.

Systematic uncertainties on $\omega_a$ are estimated as follows.
A timing shift due to pile-up of hits in the tracking detector is estimated as less than 36~ppb
in the detector simulation by taking into account time responses of readout electronics.
A correction for a pitch angle is not necessary in the case of muon storage 
in a perfect weak magnetic focusing field~\cite{Semertzidis:2016kte}. 
A difference in the actual field distribution 
from the perfect case leads to a systematic uncertainty of 13~ppb which is estimated from a precision spin-tracking simulation
of muon beam storage.
Residual electric fields modify $\omega_a$ through the $\vec{\beta} \times \vec{E}$ term.
With 1~mV/cm monitoring resolution for an E-field, the error on $\omega_a$ is 10~ppb. 
Other effects, such as distortion of the time distribution due to high-energy positrons 
hitting the detector at delayed timing and differential decay due to the momentum spread of the muon beam, 
are of the order 1~ppb.
 In the $\omega_p$ measurement, absolute calibration of the standard probe has an uncertainty of 25~ppb.
Positioning resolution of the field mapping probe at the calibration point and the muon storage
region leads to 20~ppb and 45~ppb uncertainties, respectively.
Other effects, such as field decay and eddy currents from the kicker are less than 10~ppb.
Table~\ref{tab:systematics} summarizes systematic uncertainties on $a_{\mu}$.
We estimate that the combined systematic uncertainty on $a_{\mu}$ is less than 70~ppb.

\begin{table}[t]
  \caption{Summary of statistics and uncertainties}
  \label{tab:sensitivity}
  \begin{center}
  \begin{tabular}{|p{0.5\textwidth}|c|c}
  \hline
          & Estimation \\
    \hline
    \hline
    Total number of muons in the storage magnet & $5.2 \times 10^{12}$ \\
    Total number of reconstructed $e^+$ in the energy window [200, 275~MeV] &  $5.7\times 10^{11}$ \\
    Effective analyzing power &  0.42 \\
    Statistical uncertainty on $\omega_{a}$ [ppb]  &  450 \\
    \hline
    Uncertainties on $a_{\mu}$ [ppb]  & 450 (stat.) \\
                                                            & $<70$ (syst.) \\
    Uncertainties on EDM [$10^{-21}~e\cdot$cm]  & 1.5 (stat.) \\
                                                                              & 0.36 (syst.) \\
    \hline
  \end{tabular}
  \end{center}
\end{table}

\begin{table}[t]
  \centering
  \caption{Estimated systmatic uncertainties on $a_{\mu}$}
    \begin{tabular}{|  p{3.0cm}   r||  p{5.0cm}   r|}
    \hline
    \multicolumn{2}{|c||}{Anomalous spin presession ($\omega_a$)} & \multicolumn{2}{c|}{Magnetic field ($\omega_p$)}\\
    Source & Estimation (ppb)  & Source & Estimation (ppb) \\
    \hline
    \hline
    Timing shift       & $<36$ & Absolute calibration & 25\\ 
    Pitch effect       &    13 & Calibration of mapping probe& 20\\ 
    Electric field     &    10 & Position of mapping probe& 45\\ 
    Delayed positrons  &    0.8 & Field decay & $<10$\\ 
    Diffential decay   &    1.5 & Eddy current from kicker& 0.1 \\ 
    \hline
    Quadratic sum      &  $<40$ & Quadratic sum & 56 \\ 
    \hline
    \end{tabular}
  \label{tab:systematics}
\end{table}

 A muon EDM will produce muon spin precession out of
the horizontal plane that is defined by the ideal muon orbit.
This can be seen from Eq.~(\ref{eq:omega_tot})
where the second term is the EDM term that is perpendicular
to the $a_\mu$ term. Due to the fact that the EDM term generates
vertical motion of the spin, one can extract the EDM term
from the oscillation of the up and down asymmetry $\mathcal{A}_{\text{UD}}(t)$ in
the number of positrons detected,
\begin{eqnarray}
\mathcal{A}_{\text{UD}}(t) = 
\frac{N^{\text{up}}(t) - N^{\text{down}}(t)}{N^{\text{up}}(t) + N^{\text{down}}(t)} =
\frac{PA_{\text{EDM}} \sin{(\omega t+\phi)}}{1+ P A \cos{(\omega t+\phi)}},
\end{eqnarray}
where $P$, $A$, and $\phi$ are the polarization of the muon and an effective analyzing power of muon decay, 
and a phase of muon spin with respect to direction of the momentum,
respectively. $A_{\rm EDM}$ is an effective analyzing power associated with the EDM.
A simulated up-down asymmetry in the case of $d_{\mu}=1\times 10^{-20}~e\cdot\mbox{cm}$ 
is shown in Fig.~\ref{fig:Wiggle}~(right).
The estimated statistical sensitivity for EDM is $1.5\times 10^{-21}~e\cdot\mbox{cm}$ (See Table~\ref{tab:sensitivity}).

A major source of systematic uncertainty on EDM is detector misalignment with respect to
the plane of the muon storage. The alignment resolution is estimated as 3.6~$\mu$rad~\cite{YasudaMT}
from the resolution 
of the alignment monitor system made with optical frequency comb technology.
This leads to the systematic uncertainty of $0.36\times 10^{-21}~e\cdot\mbox{cm}$.
Effects of axial electric field and radial magnetic field~\cite{Silenko:2017vvd}
are both less than $10^{-24}~e\cdot\mbox{cm}$, thus negligibly small.

\section{Summary}\label{sec:Summary} 

\rev{A} new method of measuring $a_{\mu}$ and EDM of the muon is described.
\response{Our} experiment utilizes a low-emittance muon beam prepared by 
reaccelerating thermal-energy muons created from laser-resonant 
ionization of muonium atoms. The low emittance muon beam allows use of very
weak magnetic focusing and the selected low muon momentum (300~MeV/$c$) leads to the use of a compact magnetic storage ring, 
instead of the strong electric focusing at the magic momentum (3~GeV/$c$) used by the previous and ongoing $g-2$ experiments.
A novel three-dimensional spiral injection method with a pulsed magnetic kick
is adopted to store the muon beam in the storage ring \rev{efficiently}.
\response{Our} experiment reconstructs positron tracks from muons decaying during their storage 
with a tracking detector consisting of silicon-strip sensors.

\response{Our} experiment intends to reach statistical
uncertainties for $a_{\mu}$ of 450~ppb and for muon EDM of
$1.5\times 10^{-21}~e\cdot\mbox{cm}$, for an acquisition time
of $2.2 \times 10^7$ seconds. The statistical precision is comparable to
that of the BNL experiment. The EDM sensitivity is about two orders of magnitude
higher than the BNL limit.
{Present estimates of systematic uncertainties on $a_{\mu}$ and EDM are
factors of seven and four smaller than the statistical uncertainties, respectively.
\response{Our} experiment with statistically limited sensitivity 
will test the 3\,$\sigma$ deviation on $g-2$ reported by
the BNL experiment with significantly different and improved systematic uncertainties 
and will search for new sources of T-violation in the muon EDM with unprecedented sensitivity.

\section*{Acknowledgments}
The authors would like to thank the KEK and the J-PARC muon section staffs for their strong support.
This work is supported by JSPS KAKENHI Grants No.~JP19740158, No.~JP23108001, No.~JP23740216, No.~JP25800164, No.~JP26287053, No.~JP26287055, No.~ JP15H03666, No.~JP15H05742, No.~JP16H03987, No.~JP16J07784, No.~JP16K13810, No.~JP16K05323, No.~JP17H01133, No.~JP17H02904, No.~JP17K05466, No.~JP17K18784, No.~JP18H01239 and No.~JP18H03707. 
This work is also supported by the Korean National Research Foundation Grants No.~NRF- 2015H1A2A1030275, No.~NRF-2015K2A2A4000092, and No. NRF-2017R1A2B3007018; 
the Russian Foundation for Basic Research Grant No.~RFBR 17-52-50064 which is a part of the Japan-Russia Research Cooperative Program; the Russian Science Foundation Grant No. 17-12-01036; the Russian
Ministry of Science and Higher Education Agreement 14.W03.31.0026; 
the U.S.-Japan Science and Technology Cooperation Program in High Energy Physics;
the Discovery Grants Program of the Natural Sciences and Engineering Research Council of Canada;
and the Institute for Basic Science (IBS) of Republic of Korea under Project No. IBS-R017-D1-2018-a00.




\bibliographystyle{ptephy}

%
%
%
%
\end{document}